\documentclass{ieeeaccess}
\usepackage{cite}
\usepackage{amsmath,amssymb,amsfonts}
\usepackage{algorithmic}
\usepackage{graphicx}
\usepackage{textcomp}
\usepackage{ulem}
\usepackage[printonlyused,smaller]{acronym}
\usepackage[table]{xcolor} 
\usepackage{pifont}
\usepackage{multirow}
\usepackage{booktabs} 
\newcommand{\cmark}{\ding{51}} 
\newcommand{\xmark}{\ding{55}} 
\usepackage{array}
\usepackage{url}

\def\BibTeX{{\rm B\kern-.05em{\sc i\kern-.025em b}\kern-.08em
    T\kern-.1667em\lower.7ex\hbox{E}\kern-.125emX}}
\begin{document}
\history{Date of publication xxxx 00, 0000, date of current version xxxx 00, 0000.}
\doi{10.1109/ACCESS.2017.DOI}

\title{Energy Efficiency in Network Slicing: Survey and Taxonomy}
\author{\uppercase{Adnei Willian Donatti}\authorrefmark{1}, \uppercase{Marcia Cristina Machado}\authorrefmark{1}, \uppercase{Marvin Alexander Lopez Martinez}\authorrefmark{2}, \uppercase{Sabino Rogério S. Antunes}\authorrefmark{2}, \uppercase{Eli Carlos Figueiredo Souza}\authorrefmark{2}, \uppercase{Sand Correa}\authorrefmark{3}, \uppercase{Tiago Ferreto}\authorrefmark{4}, \IEEEmembership{Member, IEEE}, \uppercase{José Augusto Suruagy}\authorrefmark{2,5}, \IEEEmembership{Life Senior Member, IEEE}, \uppercase{Joberto S. B. Martins}\authorrefmark{6}, \IEEEmembership{Life Senior Member, IEEE}, and \uppercase{Tereza Cristina Carvalho}\authorrefmark{1}}
\address[1]{Universidade de São Paulo (USP), São Paulo, SP 05508-090, Brazil (e-mail: adnei.donatti@usp.br, macrismachado@usp.br, terezacarvalho@usp.br)}
\address[2]{Universidade Federal de Pernambuco (UFPE), Recife, PE 50740-560, Brazil (e-mail: malm@cin.ufpe.br, ecfs@cin.ufpe.br, sabino.antunes@ufpe.br, suruagy@cin.ufpe.br)}
\address[3]{Universidade Federal de Goiás (UFG), Goiânia, GO 74690-900, Brazil (e-mail: sand@inf.ufg.br)}
\address[4]{Pontifícia Universidade Católica do Rio Grande do Sul (PUCRS), Porto Alegre, RS 90619-900, Brazil (e-mail: tiago.ferreto@pucrs.br)}
\address[5]{CESAR.School, Recife, PE 50030-220, Brazil (e-mail: suruagy@cesar.school)}
\address[6]{Universidade Salvador (UNIFACS), Salvador, BA 41770-235, Brazil (e-mail: joberto.martins@animaeducacao.com.br)}

\tfootnote{The authors thank FAPESP/MCTIC/CGI cooperation agreement under the thematic research project 2018/23097-3 - SFI2 - Slicing Future Internet Infrastructures, Instituto ÂNIMA (IA), and the National Council for Scientific and Technological Development (CNPq).}

\markboth
{Donatti \headeretal: Energy Efficiency in Network Slicing: Survey and Taxonomy}
{Donatti \headeretal: Energy Efficiency in Network Slicing: Survey and Taxonomy}

\corresp{Corresponding author: Adnei Donatti (e-mail: adnei.donatti@usp.br).}

\begin{abstract}
\acf{NS} is a fundamental feature of 5G, 6G, and future mobile networks, enabling logically isolated virtual networks over shared infrastructure. As data demand increases and services diversify, ensuring \ac{EE} in \ac{NS} is vital (not only for operational cost savings but also to reduce the \ac{ICT} sector’s environmental footprint). This survey addresses the need for a comprehensive and holistic perspective on energy-efficient \ac{NS} by reviewing and classifying recent strategies across the \ac{NS} life cycle. Our contributions are threefold: (i) a thorough review of state-of-the-art techniques aimed at reducing energy consumption in \ac{NS}; (ii) a novel taxonomy that organizes strategies into infrastructure, path/route, and slice operation levels; and (iii) the identification of open challenges and research directions, with a focus on systemic, cross-layer, and AI-driven approaches. By consolidating insights from recent developments, our work bridges existing gaps in the literature, offering a structured foundation for researchers and practitioners to design, evaluate, and improve energy-efficient network slicing systems.
\end{abstract}

\begin{keywords}
Artificial Intelligence, Energy-Efficient Slicing, Energy-Efficient Slicing Strategy, Energy Efficiency, Network Slicing, Taxonomy.
\end{keywords}

\titlepgskip=-15pt

\maketitle

\section{Introduction}
\label{sec:introduction}

\PARstart{E}{nergy} consumption and its environmental consequences are pressing concerns for modern society, particularly in the design of computer and telecommunication systems. In quantitative terms, the \ac{CT} sector was responsible for an estimated 1.8\% to 3.9\% of global \ac{GHG} emissions in 2020~\cite{freitag_2021}. A more alarming perspective from Andrae~\cite{huawei_andrae_update_2020, huawei_andrae_2015} suggests that this figure may have been as high as 6.3\% in 2020, with projections indicating a potential increase to 23\% by 2030.

While the use of renewable energy sources offers a promising path to reducing environmental impact, challenges related to their supply, distribution, and scalability remain significant~\cite{huawei_andrae_update_2020}. Consequently, given the current global energy generation matrix (still heavily reliant on fossil fuels), energy consumption from computer systems is expected to continue contributing substantially to environmental degradation~\cite{lorincz_greener_2019}.

In this context, \acl{NS} emerges as a promising paradigm that not only addresses performance and scalability demands but also opens opportunities for energy efficiency. By enabling resource sharing and dynamic allocation, \ac{NS} can significantly contribute to reducing overall energy consumption.

\acl{NS} is a technological enabler that allows the sharing of physical infrastructure among multiple virtualized network instances, promoting high resource utilization and operational efficiency. Widely adopted in computer systems and modern telecommunication networks (especially in 5G and beyond deployments), \ac{NS} plays a pivotal role in improving energy optimization. Furthermore, it is expected to be a cornerstone in future 6G networks, which demand advanced virtualization, adaptive flexibility, and intelligent resource orchestration~\cite{martins_enhancing_2023}.

Achieving high \ac{EE} in \acl{NS} means addressing a global environmental and sustainability concern and, no less important, reducing costs for the computer and telecommunication sectors.

In this context, several energy-efficient methods, algorithms, and strategies for computer systems and telecommunications have been studied. However, a gap exists in addressing energy efficiency in \ac{NS}. Motivated by this, this work surveys and proposes a taxonomy to contribute to the community's efforts to promote \ac{EE} in Network Slicing.

This study aims to answer the following question: ``How can one contribute to increasing energy efficiency in network slicing?". To answer this question, the paper identifies the state-of-the-art of energy efficiency in \ac{NS} and proposes a taxonomy to classify the strategy and methods identified.

The main contributions of this study can be summarized as follows:
\begin{itemize}
    \item A thorough review of state-of-the-art techniques aimed at reducing energy consumption in \ac{NS}; 
    \item A novel taxonomy that organizes strategies into infrastructure, path/route, and slice operation levels; and
    \item The identification of open challenges and research directions, with a focus on systemic, cross-layer, and \acs{AI}-driven approaches.
\end{itemize}

\begin{table*}[!ht]
\centering
\caption{Related work comparison.} \label{tab:related_surveys}
\renewcommand{\arraystretch}{1.3} 
\setlength{\arrayrulewidth}{0.8pt} 
\rowcolors{2}{gray!20}{white} 
\begin{tabular}{|
    >{\centering\arraybackslash}m{1.5cm} |
    >{\centering\arraybackslash}m{3cm} |
    >{\centering\arraybackslash}m{3cm} |
    >{\centering\arraybackslash}m{3cm} |
    >{\centering\arraybackslash}m{1.5cm} |
    >{\centering\arraybackslash}m{2.5cm} |}
    \hline
    \textbf{Ref.} & \textbf{Survey Target} & \textbf{EE Techniques Review} & \textbf{Taxonomy — NS-specific EE Techniques} & \textbf{NS Review} & \textbf{NS lifecycle oriented} \\
    \hline

    \cite{alamu_survey_techniques_hetnets_2020} & HetNets & \textcolor{blue}{\cmark} & \textcolor{red}{\xmark} & \textcolor{red}{\xmark} & \textcolor{red}{\xmark} \\
    \cite{buzzi_survey_ee_techniques_5G_2016} & 5G Networks & \textcolor{blue}{\cmark} & \textcolor{red}{\xmark} & \textcolor{red}{\xmark} & \textcolor{red}{\xmark} \\
    \cite{larsen_toward_2023} & 5G and Beyond RAN & \textcolor{blue}{\cmark} & \textcolor{red}{\xmark} & \textcolor{red}{\xmark} & \textcolor{red}{\xmark} \\
    \cite{setiawan_energy-efficient_2024} & Softwarized Networks & \textcolor{blue}{\cmark} & \textcolor{red}{\xmark} & \textcolor{blue}{\cmark} & \textcolor{red}{\xmark} \\
    \hline

    \textbf{This work} & \textbf{E2E Network Slicing} & \textcolor{blue}{\cmark} & \textcolor{blue}{\cmark} & \textcolor{blue}{\cmark} & \textcolor{blue}{\cmark} \\
    \hline
\end{tabular}
\end{table*}

The sections of this paper are structured as follows. Section~\ref{sec:related_work} presents the related work. Section~\ref{sec:background} provides the reader with some background, reviewing \ac{NS} fundamental concepts and associating them with the energy efficiency scenario. Section~\ref{sec:strategies} identifies the strategies used to increase \ac{EE} in \ac{NS}. Section~\ref{sec:taxonomy} presents the new taxonomy for \ac{EE} methods in \ac{NS} and maps the surveyed work on this taxonomy. Section~\ref{sec:trends_challenges} discusses the taxonomy and possible research directions and, finally, Section~\ref{sec:conclusion} presents the final considerations.

\section{Related Work}\label{sec:related_work}

Although many papers have been written regarding \ac{EE} in 5G networks, fewer of them consider \ac{NS}. In fact, up to the submission of this paper, the classification of methods and techniques for energy efficiency in network slicing is often addressed in a generalist fashion by research topics or technology, for example, resource allocation, network planning, and energy harvesting. 
However, we want to provide a taxonomy based on the adopted strategies for achieving energy efficiency gains, classifying them according to the optimization level in which they apply: infrastructure, path/route, and slice operation.
Existing works \cite{alamu_survey_techniques_hetnets_2020, buzzi_survey_ee_techniques_5G_2016, larsen_toward_2023, setiawan_energy-efficient_2024} are relevant to our proposal and are further presented in Table~\ref{tab:related_surveys}. 
We compare them to our work according to the survey's target, the review of \ac{EE} techniques, the proposal of a taxonomy including specific aspects of \ac{NS} with \ac{EE} method, fundamentals review on \ac{NS}, and \textcolor{black} the \ac{3GPP} \ac{NS} life cycle.

The authors in~\cite{alamu_survey_techniques_hetnets_2020} cover \ac{EE} techniques in ultra-dense \acp{HetNet}.
The overview of \ac{EE} techniques lists five categories: \textbf{1.} Network Planning and Deployment; \textbf{2.} Optimization of Radio Transmission Process; \textbf{3.} Base Station Sleeping Strategy; \textbf{4.} Hardware Solution; and \textbf{5.} Energy Harvesting and Transfer.
The document structure presents every category with techniques and their respective publications. 
The authors also discuss the problem formulation of \ac{EE} in \acp{HetNet}, presenting results in the literature for power consumption models and energy efficiency metrics, just for a \ac{BS} without any comment about functional
splitting nor \ac{NS}.
The work ends by discussing future directions and lessons learned, contributing to summarizing the scenario.
Our proposal is similar to that of\cite{alamu_survey_techniques_hetnets_2020}, as we want to investigate \ac{EE} techniques for \ac{NS}, which can be applied to \acp{HetNet}. 
Even though it presents recent works on \ac{EE} techniques in wireless mobile networks, and some physical techniques such as sleep mode and Radio Frequency (RF)
transmission which are also included in our taxonomy, their contributions lack investigation on the \ac{NS} technique, a cornerstone for future networks.

\Figure[!htb](topskip=0pt, botskip=0pt, midskip=0pt)[width=\textwidth]{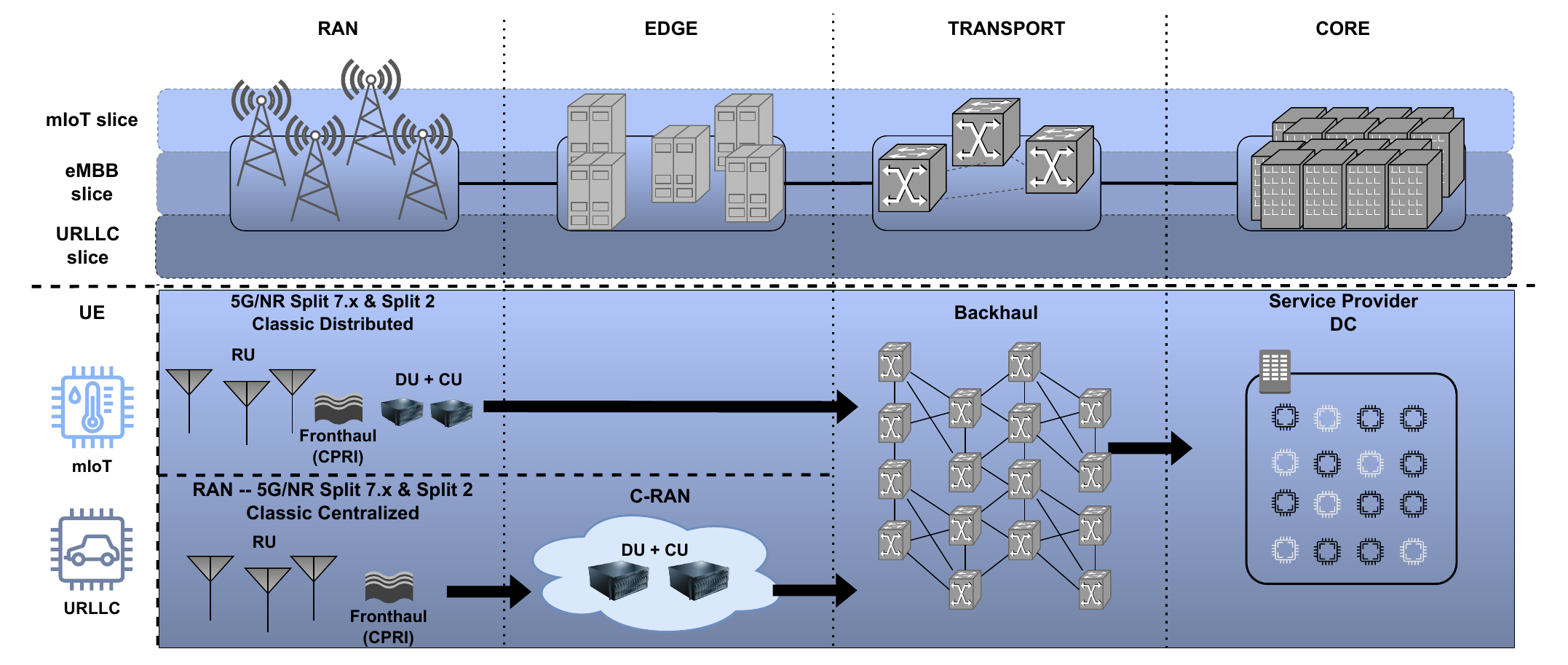}
{Network Slicing in a 5G scenario.\label{fig:5G_VNFs}}

The survey presented in \cite{buzzi_survey_ee_techniques_5G_2016} was published in 2016 when 5G was still in its beginnings.
The authors propose a taxonomy for energy-efficient techniques for 5G networks grouping them into four broad categories: \textbf{1.} Resource Allocation; \textbf{2.} Deployment \& Planning; \textbf{3.} Hardware Solutions; and \textbf{4.} Energy Harvesting \& Transfer.
Back then, the authors pointed out the need for a holistic approach combining multiple techniques.
They included in hardware solutions the cloud and edge based implementation of a \ac{RAN} with \ac{NFV}, but
without any mention about \ac{NS}.
In our research, the evolution in this field is clear. 
We have identified energy-efficient techniques based on slice orchestration, considering an \ac{E2E} view of all deployed slices.
Furthermore, given the evolution of research in the context of \ac{NS} in 5G, we propose a novel taxonomy classifying techniques according to their strategy to increase energy efficiency. 
We also summarize and represent techniques according to the optimization level they achieve: \textbf{1.} Slice Usage Optimization; \textbf{2.} Path/Route Optimization; and \textbf{3.} Infrastructure Optimization.

In \cite{larsen_toward_2023}, the authors present a study on energy efficiency for 5G/6G \ac{RAN}. 
Their research focuses on RAN network architecture, technology evolution, and network sharing. 
In the network architecture topic, they discuss the importance of the \acp{VNF} location of processing versus its impact on performance parameters such as delay in task execution and data transmission. 
Considering energy efficiency in \ac{RAN}, they discuss three main architectural approaches: \ac{vRAN}, \ac{C-RAN}, or \ac{O-RAN}. 
\ac{vRAN} supposes that \acp{VNF} can be run in any hardware and software components that are more energy efficient. 
C-RAN is a more centralized option. Energy efficiency is achieved by concentrating processing in a central location, which can result in a loss of network performance.
Finally, \ac{O-RAN} is based on open interfaces in mobile networks, allowing the adoption of multi-vendor solutions that consume the least amount of energy. 
In the case of technological evolution, the main technologies involve improvements in the power amplifier, spectral efficiency, reduced signaling, sleep modes, network virtualization, and \ac{AI}.a

They also present several energy improvements enabled by 6G, such as zero touch networks, rate splitting, intelligent reflecting surfaces, improved sampling techniques, and energy harvesting.
In the sequence, they briefly present methods for achieving CO2 neutrality.

Finally, they discuss network sharing, which includes: roaming; sharing the entire  RAN (except radio frequencies) between multiple operators; sharing of the RAN as a whole (including radio frequencies) among multiple core networks; and \ac{NS}. 
At the end, the paper provides guidelines for an operator to move towards greener networks and a map of the different \ac{RAN} energy-savings possibilities versus consumer experience impact, considering three levels of difficulty in implementing these approaches (minimum, medium, and maximum effort).

The survey carried out by Setiawan et al. \cite{setiawan_energy-efficient_2024} covers \ac{EE} in \ac{SDN}, \ac{NFV}, and \ac{NS}, classifying works according to their network segment (that is, data center, transport, wireless, and emerging: IoT, satellite, vehicular, etc.), evaluation method, metrics (\textit{e.g.}, capacity, latency), and layer.
The authors fit \ac{EE} strategies from the three different subjects (\ac{SDN}, \ac{NFV}, and \ac{NS}) into six categories: 1. Hardware-based improvements; 2. Dynamic Adaptation (DA); 3. Sleep Modes (SM); 4. Heterogeneous Network (HT); 5. Energy Harvesting (EH); and 6. Machine Learning (ML).
Sharing these categories among \ac{SDN}, \ac{NFV}, and \ac{NS} results in a broad general classification, giving up details that are explored by each \ac{EE} mechanism according to the specific subject.
For example, approaches that rely on Dynamic Adaptation, Machine Learning, or Sleep Mode may work differently depending on if they are applied to \ac{NS} or \ac{SDN}.
They conclude their paper discussing future research challenges that range from programmable hardware to experimental evaluation environments.
For each softwarized technology (including NS) they classify the works according to their network segment, while we aim to provide a taxonomy considering the specific aspects and details of \ac{NS}
strategies according to their optimization level.

Unlike previous works that offer high-level taxonomies or general discussions of \ac{EE} across mobile networks, our proposed taxonomy presents a more focused and detailed classification centered on \acf{NS}.
It organizes \ac{EE} strategies based on multiple dimensions, including optimization level (infrastructure, path/route, and slice operation) and the underlying mechanisms employed.
Additionally, our taxonomy incorporates both the network segment and the slice instance life cycle, providing alignment with end-to-end management frameworks as defined in 3GPP-compliant \ac{NS} deployments.
By doing so, our work offers a clearer and more targeted understanding of how \ac{EE} is approached specifically within network slicing (a perspective not thoroughly explored in prior studies).

\section{Background} \label{sec:background}

\subsection{Network Slicing Fundamentals} \label{subsec:ns_fundamentals}

The concept of \textit{Network Slicing} was introduced in 2015 by the \ac{NGMN} Alliance \cite{5G-NS-NGMN,5G-NS-VTC}.
According to~\cite{3gpp_2023_ts28530_18_0}, a network slice is ``\textit{a logical network that provides specific network capabilities and network characteristics, supporting various service properties for network slice customers}''.
The network slicing technology is based on ``slicing'' the physical network into several virtual networks.
Each of these virtual networks has its own requirements (\textit{e.g.}, latency, throughput, security) granted by employing \ac{VNF} chains.
In addition, the combination of \ac{SDN} and \acp{VNF} makes it possible to achieve better flexibility and management of the virtualized network and services, benefiting from network programmability, flow forwarding, lower cost, elasticity, and load balancing~\cite{slice_how_to_2019, nso_survey}.
Furthermore, these virtualization techniques allow network operators to move from specialized hardware and software to cloud-based solutions. 
For example, in 5G mobile networks, the \ac{RAN} segment can be split so that some functions previously associated to the \ac{BS} are now virtualized and hosted in a \ac{C-RAN}. 
Figure~\ref{fig:5G_VNFs} illustrates three slices with different requirements, \ac{mIoT}, \ac{eMBB}, and \ac{URLLC}, sharing the same underlying physical infrastructure.
This figure also shows how the \ac{DU} and \ac{CU} functions (traditionally associated to the Base station) become virtual functions inside \ac{C-RAN}.

Most \acp{NSP} or \acp{ISP} take on the role of \acp{SP}, since these operators already own the infrastructure.
In this sense, \acp{SP} can also provide \ac{NSaaS} depending on the customer demand.
Organizations such as ETSI-3GPP and 5GPPP play an important role in standardization, providing definitions and recommendations for service providers~\cite{3gpp_2023_ts28530_18_0, 5g_ppp_5g_vision}.
In this sense, the slice provisioning workflow is separated into three phases, the network slice life cycle (Figure~\ref{fig:slice_instance_life_cycle})~\cite{3gpp_2023_ts28530_18_0}:

\begin{itemize}
    \item \textbf{Preparation}: During this phase, the network slice is prepared for instantiation. 
    This phase includes tasks such as slice design, capacity planning, and network function evaluation.
    \item \textbf{Commissioning}: This phase deals with the provisioning of the network slice instance. 
    The main tasks that take place in this phase are resource allocation and configuration to meet \ac{QoS} requirements, which can lead to the creation or modification of existing instances;
    \item \textbf{Operation}: This phase begins with the activation of the slice instance.
    The operation phase also includes tasks related to supervision, monitoring of \acp{KPI}, capacity planning, modification, and deactivation of slice instances; and
    \item \textbf{Decommissioning}: This phase handles tasks such as releasing previously allocated resources, removing configurations, and deleting an instance.
\end{itemize}

\Figure[!htb](topskip=0pt, botskip=0pt, midskip=0pt)[width=0.99\columnwidth]{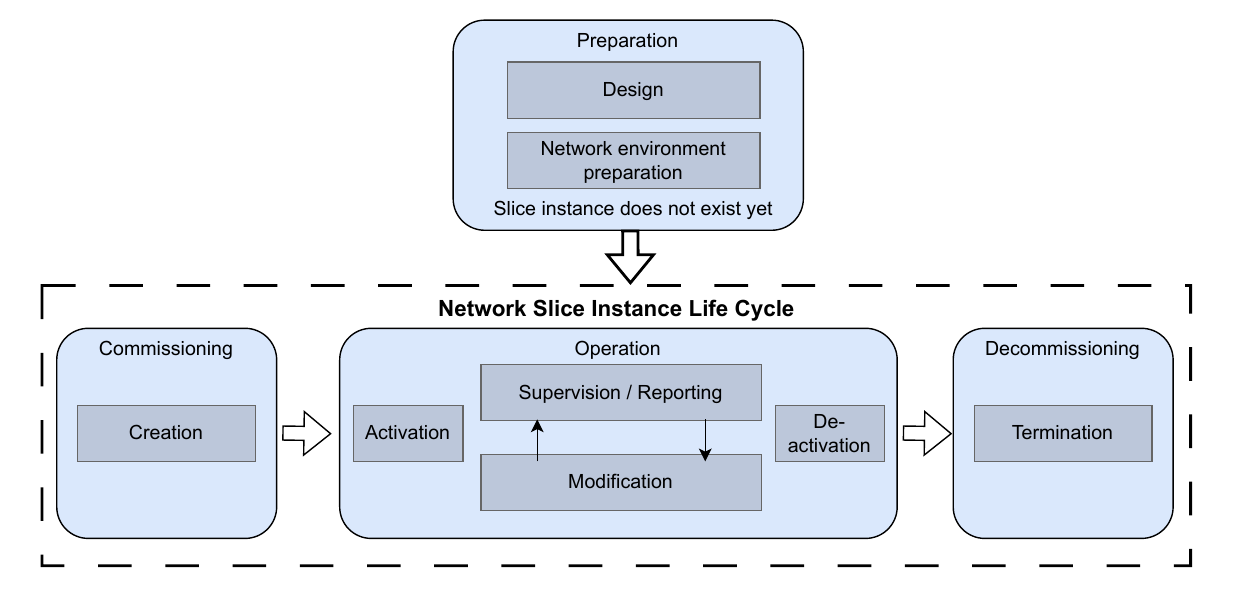}
{\textcolor{blue}{Network slicing} instance life cycle.\label{fig:slice_instance_life_cycle}}

\Figure[!htb](topskip=0pt, botskip=0pt, midskip=0pt)[width=0.99\columnwidth]{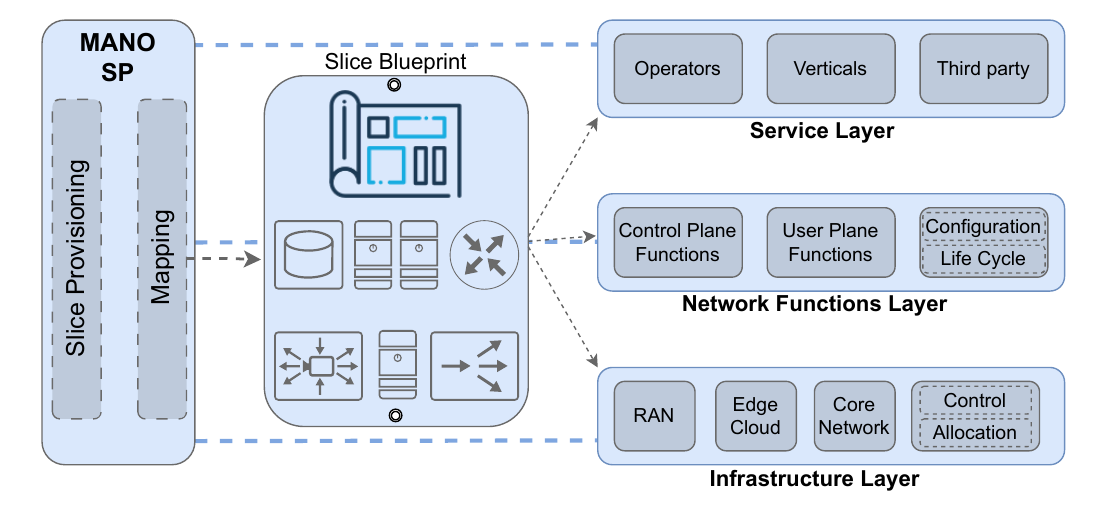}
{Generic Framework representing basic elements and layers in Network Slicing (adapted from~\cite{foukas_survey_ns_2017}).\label{fig:arch_vision}}

According to~\cite{foukas_survey_ns_2017}, the architectural vision of \ac{NS} applied to 5G can be mapped onto a generic framework composed of infrastructure, network functions, and service layers.  
In addition, a \ac{MANO} entity manages and coordinates the elements inside each layer (Figure~\ref{fig:arch_vision}). 

In the framework depicted in Figure~\ref{fig:arch_vision}, the infrastructure layer comprises the physical network infrastructure (e.g., routers, cables, switches), including its deployment, control, and management (e.g., allocation of physical resources). 
The Network Functions Layer is concerned with the configuration and life cycle management of VNFs, which are placed on top of the virtual infrastructure created and chained together to offer an E2E service that meets all the specified requirements. 
Finally,  the Service Layer, which distinguishes Network Slicing from other forms of slicing that have existed in the past (mainly in testbeds and cloud environments), comprises the E2E service description that will be mapped onto a blueprint of all resources and \acp{VNF} needed for the network slice deployment. 
Traditionally, the service description reflects all the business requirements through an \ac{SLA} (e.g., availability, latency, and throughput)~\cite{foukas_survey_ns_2017}.

\subsection{Energy Efficiency and Machine Learning in Network Slicing}\label{subsec:ee_ml_in_ns}

\ac{NS} can potentially reshape the conventional deployment, orchestration, and operation of traditional services, even in global communication network initiatives.
However, as noted in \cite{donatti_survey_2023} and \cite{ml_ns_survey}, the complexity of the scenario pressures for solutions based on \ac{ML} and, even broadly, \ac{AI}.
In this sense, \ac{AI} becomes an essential tool in the slicing process due to its inherent ability to provide optimization and acquire knowledge~\cite{shen_ai_assisted_ns_20}, enhancing tasks such as slice orchestration~\cite{moreira_intelligent_2024}.

Allied with the ability to increase performance optimization, \ac{ML} can also address energy efficiency optimization.
Research into the use of \ac{ML} for Network Slicing \cite{donatti_survey_2023, ml_ns_survey} shows that a common strategy that seeks better energy efficiency uses \acf{FL}. 
Centralized \ac{ML} models force raw data to be transmitted across the network, which consumes more energy than spreading data processing to local computations~\cite{chegui_complementar_2021}.
In addition, some works tend to optimize energy use by applying \ac{ML} to reduce radio overhead~\cite{khan_2020, tang_2021} and CPU allocation~\cite{ayala_romero_2022, yu_nuo_2017}.

Some of the current general trends in terms of ML-based slicing include:
\begin{itemize}
    \item Embedded \ac{ML} agents in the \ac{SP} architecture \cite{moreira_enhancing_2023};
    \item Distributed learning methods such as \ac{FL} to provide architectural slicing capabilities \cite{martinez_beltran_decentralized_2023}; and
    \item Architectural data-driven approaches that support machine learning methods \cite{moreira_intelligent_2024}.
\end{itemize}

While \ac{AI}/\ac{ML} techniques enable dynamic slice optimization~\cite{donatti_survey_2023}, their deployment introduces complex energy trade-offs that must be carefully balanced.
The training phase, for instance, has a cost.
\ac{DRL} models for slice orchestration increase energy consumption significantly during learning periods due to gradient computations and parameter updates.
Even well-trained models, achieving $23$-$30$\% energy reduction in RAN slicing through predictive resource allocation, show a break-even point typically occurring after a couple of weeks of operation~\cite{ayala_romero_2022, yaser_2022}.
While \ac{FL} reduces this overhead  $10\times$ through distributed model training~\cite{blanco_luis_orchestration_2024, blanco_chergui_2022}, it still requires periodic global aggregation that consumes $3$$5$\% additional energy compared to static policies~\cite{martinez_beltran_decentralized_2023}.

Some trade-offs extend to architectural considerations in the \ac{AI}/\ac{ML} models development.
For instance, approaches relying on edge-based inference save energy by avoiding cloud transmissions~\cite{tang_2021}, but require specialized hardware accelerators whose manufacturing carbon footprint may offset operational savings~\cite{freitag_2021}.
Moreover, even the \ac{ML} model chosen has different energy proportionalities.
The authors in~\cite{thantharate_eco6g_2022} make an energy and cost analysis comparing different approaches. 
Simpler ARIMA models, for example, might reduce the energy cost by trading off final accuracy (around 15\% lower accuracy).
Finally, hybrid approaches combining heuristics for routine decisions and \ac{ML} for complex scenarios demonstrate significant efficiency while reducing computational overhead~\cite{Bolla_FITCE_2022, thantharate_eco6g_2022, blanco_luis_orchestration_2024}.

\subsection{Energy-Efficiency, Decarbonization and Sustainability}
\label{subsec:sustainability}

In our web-based, user-oriented, service-oriented, and fully connected society, energy efficiency, decarbonization, and sustainability are issues of great concern. The society concern results from the direct or indirect impact that energy consumption can have on \ac{GHG} emissions that cause climate change \cite{sundaramoorthy_energy_2023}. According to \cite{freitag_2021}, the overall global contribution of \ac{ICT} emissions is between 2.1 to 3.9\% and is concentrated in data centers, networks, and user devices. Networks are responsible for 27\% of total emissions.

Network slicing, as an important enabler for most new systems such 5G/6G, \ac{IoT}, Smart Grid, and others, contributes significantly to energy consumption worldwide. As a figure of merit, it is estimated that 5G energy consumption will account for approximately two-thirds of 1,5\% of global energy consumption. Looking at the energy consumption of 5G mobile networks, the \ac{RAN} consumes 73\% of the energy, network interconnections consume 13\%, datacenters consume 9\%, and other operations account for 5\% of energy consumption \cite{larsen_toward_2023}. These figures reveal the importance of savings in mobile networks and, above all, in the \ac{NS} as the main enabler of this technology.

\Figure[!htb](topskip=0pt, botskip=0pt, midskip=0pt)[width=0.99\columnwidth]{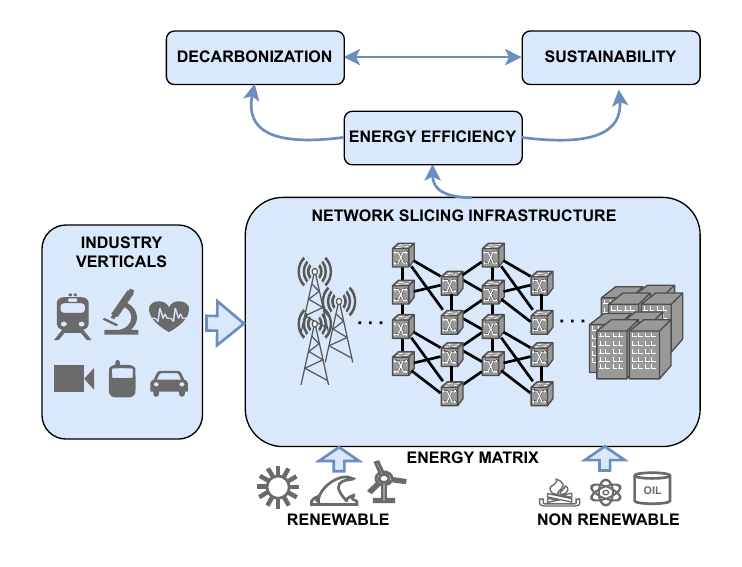}
{Correlation between \ac{NS}, energy efficiency, decarbonization, and sustainability.\label{fig:EE-Decarbonization-Sustainability}}

Regarding network slicing, there are distinct pathways to net-zero greenhouse emissions through energy efficiency techniques (Figure \ref{fig:EE-Decarbonization-Sustainability}). These pathways include enforcing energy efficiency in the processing phases of \ac{NS} and prioritizing service providers, which use renewable energy.

We try to shed some light on these alternatives by surveying and discussing how to reduce energy consumption in the network slicing process, and complementing this discussion, we clarify how energy efficiency, decarbonization, and sustainability interrelate with the network slicing process.

In the computer science domain, energy efficiency refers to the practice and methods of reducing energy consumption to perform a specific task or obtain a particular service. An energy-efficient system aims to maximize the outcomes of the tasks and services for a given amount of energy input, minimizing energy waste and guaranteeing user performance requirements (\ac{SLA}, \ac{QoS}, \ac{QoE}, and others) \cite{hafez_energy_2023} \cite{patterson_what_1996}.
Anser et al.\ \cite{anser2021energy} extend \ac{SLA} models to slices including energy-based commitments. They also propose including \ac{EE} metrics in the \ac{VNFD}.

Decarbonization refers to the ability to reduce or eliminate \ac{GHG} emissions, such as carbon dioxide (CO2) emissions, towards a net-zero pathway and is mainly associated with human activities. Decarbonization can be achieved through pathways deployed across various economic actors with a predominant focus on the industrial sector, where computer systems are important players\cite{wimbadi_decarbonization_2020}.

According to the United Nations 2030 Agenda~\cite{undesa_agenda_2030} for Sustainable Development, sustainability is a broad concept that involves three main interrelated dimensions: environmental, social, and economic. In line with this, sustainable development must ensure organizational performance while protecting people and the planet \cite{rosario_new_2023}. Alternatively, sustainable development can be conceptualized as the ability to meet the needs of the current generation without compromising the ability of future generations to meet their own needs \cite{wang_making_2021}.

Environmental sustainability refers to preserving and improving the use of natural resources with strategies that help to ensure that human activity does not compromise the Earth's resources and preserve our ecosystems. Social sustainability works to improve the well-being of the planet's citizens, addressing various issues such as equity, human rights, digital divide, and cultural diversity, among others \cite{rosario_new_2023}. Economic sustainability is related to income generation with a consequent improvement in society's quality of life. This can be achieved through technological innovation and new business models, as well as the rational use of resources.

Economic sustainability is linked to network slicing mainly by its ability to reduce carbon emissions by optimizing and reducing energy consumption in the various phases and components involved in the network slicing process. Energy-efficient network slicing also contributes to environmental sustainability with solutions and strategies in pollution control, transport systems, and production systems. When NS-based 5G and 6G technologies are used to support these systems, the contribution can be direct or indirect.

The interrelation of the concepts related to network slicing can be summarized as follows:

\begin{itemize}
    \item Decarbonization and energy efficiency are closely related concepts that establish a path toward sustainability and promote sustainable energy practices.
    \item Energy-efficient approaches to network slicing promote decarbonization and lead to a more sustainable use of energy worldwide.
    \item Another key issue for decarbonization is the deployment of clean and renewable energy sources. Countries have very different energy matrices. Currently, there is an international consensus that all countries must migrate to increasingly renewable and clean energy matrices. 
\end{itemize}

There is a balance in the energy matrix between polluting and renewable, efficient and non-polluting energy generation (Figure \ref{fig:EE-Decarbonization-Sustainability}). Although there are global commitments to increase the use of non-polluting energy, the use of polluting energy will continue to exist. In this way, energy-efficient methods for network slicing will always play an essential role, considering the two scenarios with the use of polluting and non-polluting energy, as they imply a reduction in resource consumption and a reduction in costs for the providers and users of slices.

According to \cite{freitag_2021}, the exact share of renewable energy used in the \ac{ICT} sector is unknown. Nevertheless, the sector is a significant purchaser of renewable energy, which points to a global shift towards the use of renewable energy use. In global terms, the use of renewable energy could significantly reduce the contribution of \ac{ICT}’s carbon footprint, leading to around 80\% reduction in carbon emissions.

Energy efficiency technologies, such as energy management, smart manufacturing, material efficiency, and energy efficiency methods, are well-established pathways and can potentially reduce energy use and \ac{GHG} emissions \cite{sundaramoorthy_energy_2023}. Energy efficiency is a common practice in computer science systems like \ac{IoT}, \acp{ICT}, widely used in smart cities, data centers, and initiatives such as the smart grid, which focuses on the efficient production, transmission, and distribution of energy (Figure \ref{fig:EE-Decarbonization-Sustainability}) \cite{farhan_energy_2021} \cite{hu_development_2022} \cite{jin_energy_2012} \cite{loschi_energy_2015}.

To summarize, we believe that \acf{NS} can contribute to all three pillars in sustainability:

\begin{itemize}
    \item \textbf{Environmental}: By enabling virtualized, on-demand services, NS reduces the need for always-on dedicated infrastructure, directly lowering energy usage. Efficient orchestration can reduce idle resources and facilitate green scheduling (e.g., sleep mode in underutilized RAN elements)~\cite{larsen_toward_2023, lorincz_greener_2019};
    \item \textbf{Economic}: \ac{NS} allows service providers to dynamically instantiate slices tailored to specific verticals, improving \ac{OPEX} through infrastructure sharing~\cite{xin_2017};
    \item \textbf{Social}: Tailored slices can prioritize underserved areas (\textit{e.g.}, rural 5G FWA) or critical services (e.g., healthcare), contributing to digital equity and social impact~\cite{najjuuko_rural_2021, markhasin_2017}
\end{itemize}

\section{Optimization Levels to Increase Energy Efficiency in Network Slicing}
\label{sec:strategies}

\Figure[!htb](topskip=0pt, botskip=0pt, midskip=0pt)[width=1\textwidth]{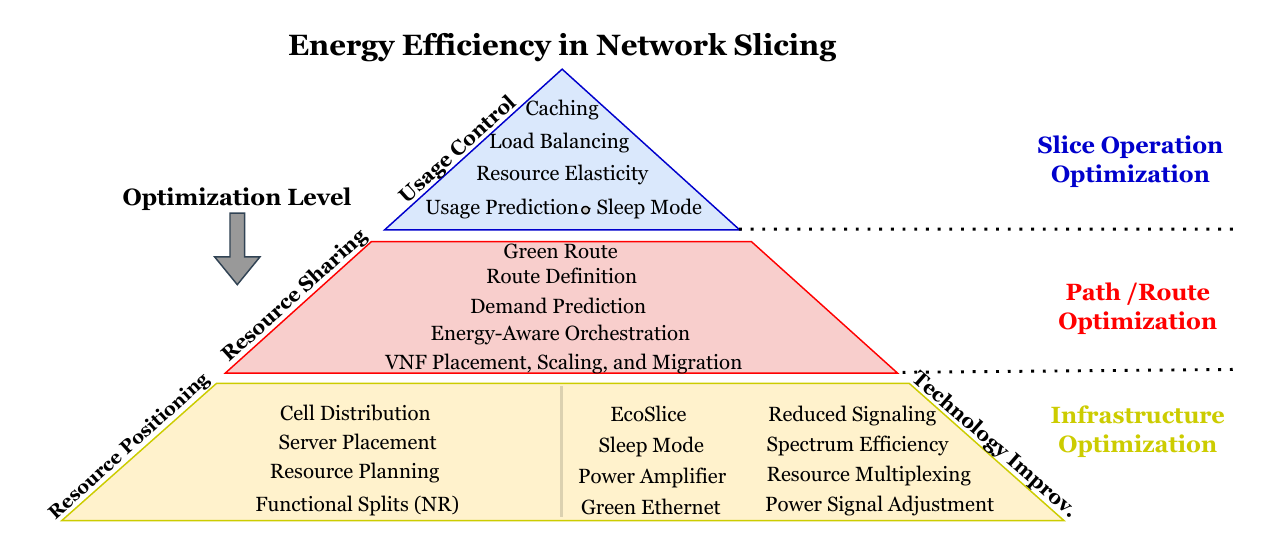}
{Optimization levels to increase energy efficiency in Network Slicing.\label{fig:pyramid_taxonomy}}

To identify the main contributions related to \acf{EE} in \acf{NS}, we defined two search queries and applied them to the IEEE and ACM Digital Libraries, covering the period from 2019 to 2024: ``Network Slicing AND Energy Efficien*'' and ``Network Slicing AND Orchestration AND Energy Efficien*''.
From the 189 papers initially retrieved, our research team filtered and selected 36 papers for in-depth discussion based on a preliminary screening of titles and abstracts.
To systematically analyze the approaches adopted across these works, we developed the taxonomy illustrated in Figure~\ref{fig:pyramid_taxonomy}, which categorizes \ac{EE} methods in \ac{NS} according to their optimization scope.
This classification emerged from recurring patterns identified during our discussions, in which the strategies to enhance \ac{EE} consistently aligned with one of three distinct levels: \textit{Infrastructure Optimization}, \textit{Path/Route Optimization}, and \textit{Slice Operation Optimization}.
Each level reflects a different scope of decision-making within \ac{NS} environments.
For instance, Infrastructure Optimization includes methods such as server placement near end-users (e.g., edge computing, \ac{BS} distribution), which reduce energy consumption through architectural design and resource positioning.
We also observed that strategies within the same level can differ in nature.
For example, while both cell distribution and spectrum efficiency techniques fall under Infrastructure Optimization, the former focuses on spatial resource allocation, whereas the latter targets technological enhancements.
Thus, our taxonomy not only organizes existing contributions but also clarifies the conceptual basis behind \ac{EE} strategies in \ac{NS}, offering a structured lens through which researchers and practitioners can evaluate and compare different approaches.

Moreover, the classification into three optimization levels emerged from our in-depth analysis and discussion of the selected literature.
Throughout the review process, our goal was to unravel the diverse approaches to increasing \ac{EE} within the context of \ac{NS}.
By systematically examining the objectives, assumptions, and methodologies of the analyzed works, we observed that \ac{EE} strategies consistently aligned with one of three distinct optimization scopes.
This three-level taxonomy effectively captures the hierarchical nature of energy-related decisions in \ac{NS} systems, ranging from structural configurations to dynamic operational controls.

Figure~\ref{fig:pyramid_taxonomy} represents the methods operating at each optimization level in the pyramid.
At the base (\textbf{Infrastructure Optimization}), methods increase energy efficiency by optimizing the infrastructure through resource positioning (e.g., functional split, cell distribution) and technology improvement strategies (e.g., sleep mode, power amplifier).
By the middle of the pyramid (\textbf{Path/Route Optimization}), techniques address energy efficiency by optimizing the slice's path/route. 
Techniques acting at this level increase energy efficiency by sharing, managing, and optimally allocating resources (resource sharing strategy). 
At the top of the pyramid (\textbf{Slice Operation Optimization}), methods work on slice usage control. 
Techniques at this level save energy by minimizing or avoiding energy waste in the slice operation context.

\subsection{Infrastructure Optimization Level}
\label{subsec:infra_decision}

In our research, we observe that methods operating at the infrastructure optimization level focus on resource positioning and technology improvements.
The architecture design of a \acf{SP} must consider energy efficiency, which, besides being environmentally friendly, also translates to \ac{OPEX} reduction.
Therefore, we consider infrastructure optimizations as the primary source of energy savings in \acf{NS}.
In this sense, the overall energy consumption is reduced by planning the computational resource capacity, positioning, and technology adopted in the slice provider infrastructure.
For instance, the computational resource capacity should be thought to increase the effectiveness of turned-on devices.
Moreover, the adopted technology in the slice provider should also corroborate lowering the energy consumption (e.g., IEEE 802.3az, Energy Efficient Ethernet).

The \ac{SP} relies on a distributed system to provide network slices. 
Therefore, the architecture design might benefit from strategies that usually derive from classical distributed systems, which precede the \ac{NS} concept and are out of the scope of this document.
For example, the publications~\cite{resource_positioning_IBM_10, dc_max_capacity_patel_13, dc_max_capacity_vishwakarma_15, energy_efficient_ethernet_22}
present contributions that are not directly related to the \ac{NS} energy efficiency (data center/cloud solutions and green Ethernet, respectively). 
However, it is noteworthy that such classical resource positioning and technology improvements could be extended to the \ac{SP} infrastructure to reduce energy consumption.

\subsection{Path/Route Optimization Level}
\label{subsec:slice_orchestration}

The Path/Route Optimization Level is based on the slice orchestration idea.
Since the term ``orchestration'' is broadly utilized in several contexts, we refer to slice orchestration as the coordination and control of the overall workflow process of tasks related to the network slice life cycle (slice preparation, commissioning, operation, and decommissioning). 
Therefore, we observe how documents address energy efficiency in coordinating and integrating multiple network slices. 
The slice provider must orchestrate all the network slices served by its infrastructure, for example, sharing resources, calculating routes, and predicting traffic peaks in routes.
Therefore, techniques operating at this level consider a panoramic view of all the slices in the system (mostly as a network graph).

\subsection{Slice Operation Optimization}
\label{subsec:usage_optimization}

Methods operating in this optimization level assume that the network slice is already in its operational phase. 
While approaches based on path/route take the provider's viewpoint (global vision of all slices), methods based on operation optimization take the slice's viewpoint. 
Therefore, most techniques optimize the usage of computational resources (e.g., computing, network, storage) and workloads. 

In Figure~\ref{fig:pyramid_taxonomy}, the strategy to increase energy efficiency using a sleep mode technique might be based on an infrastructure decision or on slice operation optimization. 
In case it is a native function from the device (\textit{e.g.}, switch, router), then we consider it a technological improvement, fitting as an infrastructure decision. 
On the other hand, if the sleep mode is asserted through an algorithm analyzing a container or a server workload, for example, the technique suits operation optimization.
Generally, works using strategies based on slice operation optimization benefit from workload and traffic fluctuation.

\section{Methods for Energy Efficiency in Network Slicing: A Taxonomy}
\label{sec:taxonomy}

In this section, we expand the pyramid presented in Figure~\ref{fig:pyramid_taxonomy} to a comprehensive taxonomy of methods for energy efficiency in \ac{NS}.
While the pyramid provides the reader with a layer-based and concise notion of such techniques structuring, we present the surveyed work alongside a detailed taxonomy depicted in Figure~\ref{fig:Taxonomy}. 
Additionally, it is noteworthy that papers in this research field might not address energy efficiency as a specific goal or measured metric. 
Instead, such documents usually address efficiency in terms of cost or resource usage, for example. 
Therefore, depending on the context, the gains in efficiency might translate directly into power usage (\textit{e.g.}, CPU and network usage).
In the following subsections, we detail surveyed work suiting every optimization level.

\Figure[!htb](topskip=0pt, botskip=0pt, midskip=0pt)[width=0.9\textwidth]{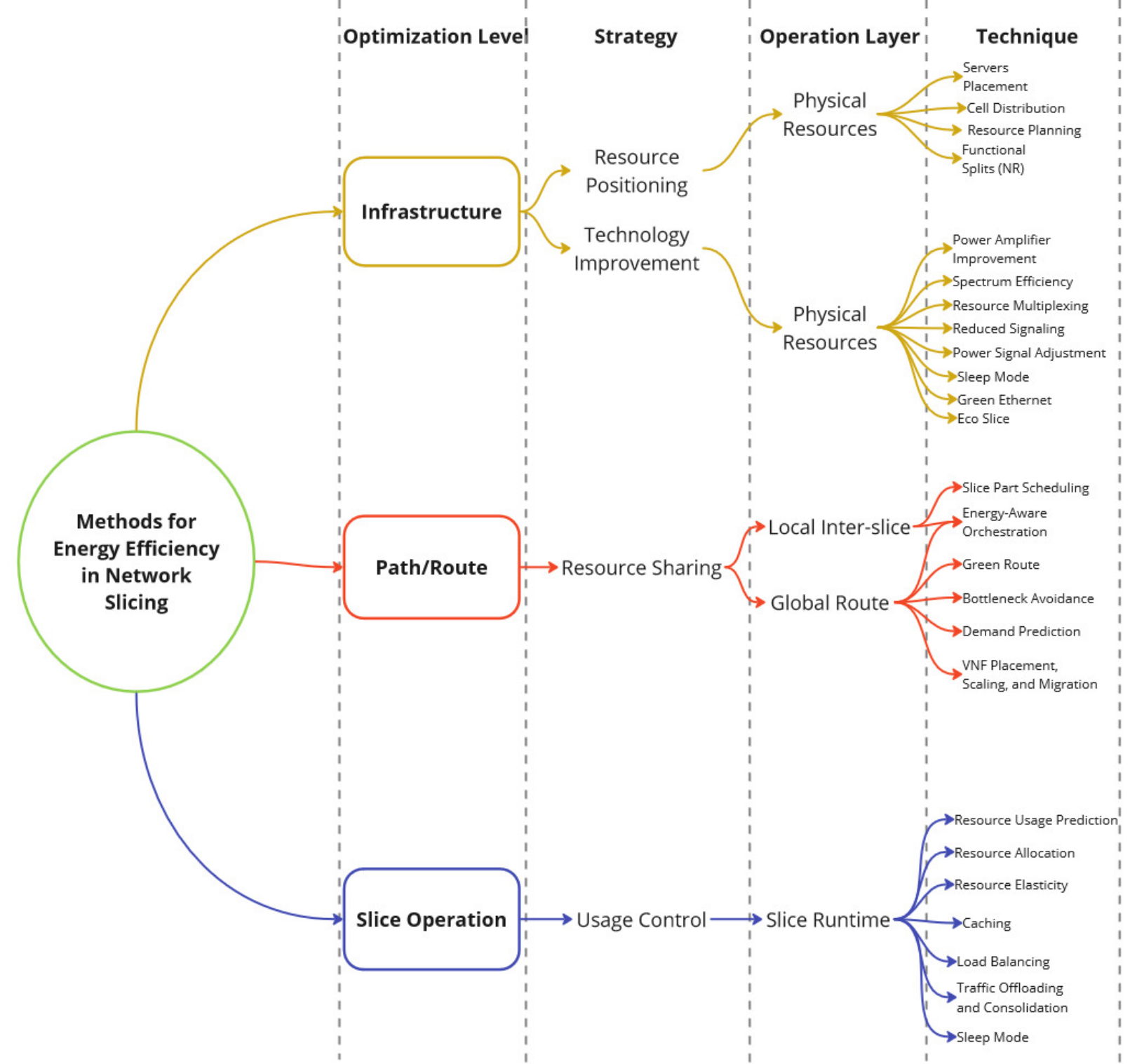}
{Taxonomy of Energy Efficiency Methods in Network Slicing.\label{fig:Taxonomy}}

\subsection{Infrastructure Optimization}
\label{subsec:opt_level_infra}

As mentioned before, \acl{NS} can benefit from traditional distributed systems \acf{EE} techniques.
Additionally, 5G \ac{NS} brings new challenges and opportunities for conventional infrastructures (mostly in the \ac{RAN} segment).
In this sense, we analyze works in the literature achieving energy efficiency through infrastructure-level optimizations.
Efforts in this category take advantage of the following strategies:

\begin{itemize}
    \item \textbf{Resource Positioning Strategy}: resource arrangement favors energy efficiency and overall performance. 
    For instance, choosing the RAN configuration, server placement, and functional splits;
    \item \textbf{Technology Improvement Strategy}: improving or adjusting technologies to reduce energy consumption. 
    For instance, power signal adjustment, resource multiplexing, and Green Ethernet.
\end{itemize}

The following tables share a common structure and summarize relevant research works that served as references to build our proposed taxonomy and its classification. Each table organizes these studies according to specific network segments, optimization levels, phases defined by the 3GPP standard, and the applied techniques, clearly classified into Artificial Intelligence (AI)-based methods, heuristic methods, and other alternative techniques.

Table~\ref{tab:infra_decision_tab} summarizes works exploring such opportunities in infrastructure-level optimization.

\begin{table*}[!ht]
\centering
\caption{Research work achieving \ac{EE} through infrastructure-level optimizations.} \label{tab:infra_decision_tab}
\renewcommand{\arraystretch}{1.3} 
\setlength{\arrayrulewidth}{0.8pt} 
\rowcolors{2}{gray!20}{white} 
\begin{tabular}{|
    >{\centering\arraybackslash}m{1cm} |
    >{\centering\arraybackslash}m{1.5cm} |
    >{\centering\arraybackslash}m{2.5cm} |
    >{\centering\arraybackslash}m{1.5cm} |
    >{\centering\arraybackslash}m{2cm} |
    >{\centering\arraybackslash}m{4cm} |}
    \hline
    \textbf{Ref.} & \textbf{Network Segment} & \textbf{3GPP Phase} & \textbf{AI Method} & \textbf{Other Methods} & \textbf{Technique} \\
    \hline

    \cite{xiang_2020_mode_selection} & F-RAN & Commissioning & \acs{RL} & --- & Resource Planning \& Positioning \\
    \cite{phyu_EE_RAN_NS_2023} & RAN & Operation & \acs{MAB} & --- & Eco Slice, Sleep Mode \\
    \cite{sen_towards_2023} & RAN & Commissioning & --- & ILP & Functional Splits (\acs{NR}) \\
    \cite{kao_qoe_sustainability_2023} & RAN & Operation & --- & \acs{LR} & Resource Planning \\
    \cite{Hossain_NOMA_2023} & RAN, Edge & Operation & --- & \acs{SACRA} & Spectrum Efficiency \\
    \cite{hossain_numerology_2023} & RAN, Edge & Commissioning & --- & \acs{MINLP} & Spectrum Efficiency \\
    \cite{ismail_advances_2020} & RAN & Operation & --- & --- & Resource Planning \\
    \cite{liu_2023} & RAN & Operation & --- & Dinkelbach, others & Spectrum Efficiency \\
    \cite{zou2022energy} & RMEC & Operation & --- & --- & Technology Improvement \\
    \cite{ndikumana_2024} & RAN, Edge Cloud & Preparation & \acs{RL} & \acs{SCA} & Functional Splits, Renewable Energy Integration \\
    \hline
\end{tabular}
\end{table*}

In the work~\cite{xiang_2020_mode_selection}, the authors explore energy efficiency in \ac{NS} by providing optimizations at the infrastructure level, taking advantage of \ac{F-RAN} to alleviate end-to-end operation (positioning strategy).
The authors investigate the impact of fog computing on power consumption and delay performance and propose a joint optimization solution of mode selection (C-RAN mode, fog radio access point mode, and device-to-device mode) and resource allocation in uplink \acp{F-RAN} for power minimization, considering predefined scenarios.
To minimize transmission costs in fronthaul and save system power, authors suggest that local data processing be enabled by \acp{F-AP} and \acp{F-UE}.
The authors use two \ac{RL} algorithms to maintain low power consumption and stable transmission delay for \acp{UE} and \acp{F-UE}.

According to the authors of~\cite{phyu_EE_RAN_NS_2023}, guaranteeing the slice requirements requires additional energy consumption.
In this sense, the authors suggest a dynamic slice activation and deactivation. 
However, the \ac{SP} architecture must contemplate a shared slice, called \textit{EcoSlice}, which operates 24/7 and provides bare-minimum service.
Thus, in conditions such as low traffic demand, operators may switch users from other slices to the \textit{EcoSlice}.
To solve the decision problem of switching users, the authors use two \ac{MAB} agents at each \ac{BS}. 
The authors claim their solution provides approximately an 11-14\% \ac{EE} improvement.
We consider the method presented by the authors as an infrastructure-level optimization since the \ac{SP} architecture must contemplate a static ``eco-aware slice'' beforehand.

In Sen and A.F.A. \cite{sen_towards_2023}, the energy efficiency is addressed by the \ac{RAN} functional split, where the function placement strategy considers different functional splits and network slice requirements. 
This is achieved through an \ac{ILP} heuristics to minimize the energy consumption of processing nodes and transport links considering the slice requirements.

In \cite{liu_2023}, the authors maximize the energy efficiency of \ac{eMBB} and \ac{URLLC} slices through the optimization of remote radio unit selection and beamforming. The problem is formulated as a \ac{MINLP}, and simulation results show that the method effectively improves \ac{EE} while assuring \ac{QoS} requirements (latency, reliability, and throughput).

Kao et al.~\cite{kao_qoe_sustainability_2023} contend that controllability of \ac{QoE} in 5G networks is a persistent issue. In response, it proposes an AI-enabled architecture designed to predict and sustain \ac{QoE}. This architecture leverages network slicing and \ac{MEC} technologies to gather cross-layer performance data in real-time and dynamically allocate network resources. Although the paper highlights the importance for telecommunications companies to adopt energy-efficient architectures not only to reduce power consumption but also to decrease the carbon footprint, it lacks detailed explanations on how the proposed architecture achieves these reductions. The paper briefly mentions that the architecture could reduce energy consumption by downscaling the number of the allocated \acp{PRB}, yet this aspect is not extensively explored.

The \ac{NOMA} technique is essential to increase spectrum efficiency in 5G cellular systems.
In this sense, Mohammad A. H. et al.\  \cite{Hossain_NOMA_2023} propose a technique for integrating \ac{NS} and \ac{NOMA} in mobile edge computing, achieving an improvement in spectral efficiency on 5G networks. They propose the \ac{SACRA} algorithm creating user clustering, computing and wireless resource allocations.
Moreover, Ismail et al.~\cite{ismail_advances_2020} review and classify recent advancements in resource allocation for 5G networks, focusing on Power-Domain \ac{NOMA} (PD-NOMA) techniques. They provide a detailed overview of the existing research into two main categories: power/energy-efficient and rate-optimal resource allocation. In terms of energy efficiency, the paper emphasizes PD-NOMA's potential to enhance power efficiency by allowing multiple users to share the same resource blocks (\textit{e.g}., time and frequency) while being separated in power levels. This method differs from traditional \ac{OMA} techniques, which allocate separate resources to each user, leading to underutilization and increased power consumption. The paper explains how PD-NOMA can make 5G networks more sustainable and use less energy by giving more power to different users and making it easier to serve many users at once.

Hossain and Ansari~\cite{hossain_numerology_2023} investigate the impact of numerology on a sliced Time Division Duplex RAN. The authors also optimize duplex ratio and power and bandwidth allocation. Moreover, they observe that, among other conclusions,  the highest numerology schemes do not necessarily translate to the highest average spectrum efficiency. 

The authors in \cite{zou2022energy} present a framework applied to \ac{RMEC}. 
In this context, network devices are energy-constrained and powered by batteries. 
The authors propose automatically generating network slices with wireless energy transfer technology to realize synchronous energy and information transmission by slices.
The authors validate their proposed framework by analyzing data from the China Telecom Sichuan Branch.

The document~\cite{ndikumana_2024} proposes a sustainable and energy-efficient \ac{O-RAN}-based architecture for deploying 5G \ac{FWA} in rural areas, where fiber deployment is costly. 
The authors introduce a three-level closed-loop system that dynamically optimizes radio resource allocation (for \ac{O-RAN} slices and Customer Premises Equipment) while minimizing energy costs. 
The edge cloud is powered by renewable energy, and the optimization process uses \ac{RL} and \ac{SCA} to balance communication utility and energy efficiency.
We consider this work into the Infrastructure Optimization category, as it primarily focuses on network infrastructure improvements (leveraging \ac{O-RAN} functional splits, renewable energy integration, and intelligent resource management to improve energy efficiency).

\subsection{Path/Route Optimization}
\label{subsec:opt_level_orchestration}

Works actuating on the route optimization level observe network slices from a global system perspective. 
The network graph is configured (and often reconfigured) to save energy. 
The strategy behind approaches at this optimization level is to distribute and share resources among different network slices sharing the same requirements. 
For instance, \ac{VNF} chains can be created based on this idea.
Furthermore, demand prediction can also be used to establish more efficient network routes.
We summarize publications achieving \ac{EE} in this optimization level in Table~\ref{tab:slice_orchestration}.

\begin{table*}[!ht]
\centering
\caption{Publications achieving \ac{EE} at path/route optimization level.} \label{tab:slice_orchestration}
\renewcommand{\arraystretch}{1.3} 
\setlength{\arrayrulewidth}{0.8pt} 
\rowcolors{2}{gray!20}{white} 
\begin{tabular}{|
    >{\centering\arraybackslash}m{1cm} |
    >{\centering\arraybackslash}m{1.5cm} |
    >{\centering\arraybackslash}m{2cm} |
    >{\centering\arraybackslash}m{2cm} |
    >{\centering\arraybackslash}m{4cm} |
    >{\centering\arraybackslash}m{4cm} |}
    \hline
    \textbf{Ref.} & \textbf{Network Segment} & \textbf{3GPP Phase} & \textbf{AI Method} & \textbf{Other Methods} & \textbf{Technique} \\
    \hline

    \cite{flexible_routing_chen_20} & Core & Commissioning & --- & \acs{MBLP} & Routing Optimization \& \ac{VNF} Placement \\
    \cite{s_joint_2023} & Edge, Core & Operation & \acs{RFR}, \acs{GBR} & Evolutionary \acs{CC} & Demand Prediction \& Slice Part Scheduling \\
    \cite{mesodiakaki20215g} & \acs{E2E} & Commissioning & --- & HERO heuristic & \ac{VNF} Placement \\
    \cite{ma_jinliang_slice_aware_2019} & RAN & Operation & --- & --- & Parametric Dinkelbach \& Route Selection \\
    \cite{xu_2022_ee_mec} & \acs{MEC} & Operation & \acs{KKT} conditions & --- & Resource Management \& Slice Part Scheduling \\
    \cite{han_ee_service_placement_mec_2022} & \acs{MEC} & Operation & --- & \acs{MILP} & Service Placement \\
    \cite{oladejo_energy-efficient_2019} & Edge, Core & Operation & --- & \acs{MILFP} & Route Optimization \\
    \cite{green_communication_chen_21} & \acs{E2E} & Full life cycle & --- & Lyapunov optimization & Network Selection \& Green Route \\
    \cite{Basu_Dynamic_2022} & Core & Commissioning & --- & \acs{MILP} & \ac{VNF} Consolidation \\
    \cite{cao_2022_RESOURCE_ALLOCATION} & Core & Commissioning & --- & Polynomial time algorithm & Slice Part Scheduling \\
    \cite{zhou2020dynamic} & Core & Commissioning, Operation & --- & Holt-Winters & Demand Prediction \\
    \cite{chen2022designing} & Core & Commissioning & \acs{RL}, \acs{QL} & --- & \ac{VNF} Placement \\
    \cite{blanco_luis_orchestration_2024} & \acs{E2E} & Operation & \acs{FL} & Constrained Neural Network & Route Optimization \& Slice Orchestration \\
    \cite{aba_ns_orchestrator_2024} & Core & Commissioning & \acs{GA} & Custom \acs{K8s} Scheduler & Slice Orchestration \\
    \hline
\end{tabular}
\end{table*}

The work presented in \cite{flexible_routing_chen_20} investigates the use of \ac{SFC} in network slicing. The authors propose minimizing the total power consumption (considering the whole network) by optimally selecting cloud nodes to deploy functions in the \ac{SFC}.
The problem is formulated as a \ac{MBLP} with \ac{E2E} latency and link capacity constraints. The authors compare their approach with other existing ones.
Following a similar logic, in~\cite{Basu_Dynamic_2022}, the authors tackle the \ac{VNF} embedding problem across \ac{SFC} instances in software-defined 5G networks. They introduce a dynamic \ac{VNF} sharing model, FlexShare-VNF, which considers flow requests for individual network slices. Implemented as a \acf{MILP} optimization, the model strategically manages \ac{VNF} sharing to enhance both network efficiency and hardware utilization. The primary aim is to boost resource utilization through sharing techniques. The authors explicitly link improved resource utilization to more energy-efficient service delivery. Although the paper asserts significant gains in energy efficiency, it primarily quantifies improvements in terms of resource utilization and latency reduction, rather than direct metrics of energy consumption.

The authors of \cite{cao_2022_RESOURCE_ALLOCATION} envision 6G networks and its requirements. They propose a novel algorithm running in polynomial time (called TailoredSlice-6G) which efficiently allocates resources within a softwarization context. The algorithm is triggered upon a slice request and then a sub-algorithm is called depending on the required resource type. The authors do not evaluate energy consumption directly. Instead, they rely on the resource consumption ratio. Results show that the proposed algorithm can be used in real networking applications.

Saibharath et al.\  \cite{s_joint_2023} devise a joint QoS and energy efficiency data-driven approach on top of pre-defined and prioritized traffic classes. 
Prioritized traffic classes are predicted using \ac{RFR} and \ac{GBR}.
The solution imposes network dynamic reconfiguration using \ac{CC} to obtain energy savings. 
Based on simulation results, authors claim node consumption is reduced through the evolutionary \ac{CC} algorithm, outperforming other standard approaches by at least 23\%.

In the paper by Oladejo and Falowo \cite{oladejo_energy-efficient_2019}, an energy-efficient \ac{MILFP} is formulated for slice resource allocation considering the \acf{QoS} of different slice use cases. The authors model a global energy efficiency method for the 5G sliced network that provides energy efficiency.

In the publication~\cite{ma_jinliang_slice_aware_2019}, the authors work on an \ac{EE} maximization problem considering bandwidth and power optimization. The approach applies to SDN \acp{HetNet}, by managing radio resources. Furthermore, the authors consider an \ac{SDN} controller with centralized control and global network information, allowing network slice selection and resource allocation to serve devices.

The publication \cite{xu_2022_ee_mec} investigates resource management methods in \ac{MEC}. The authors design numerical methods optimally allocating edge servers to reduce the average power consumption in the server.
Also exploring the benefits of \ac{MEC}, the work \cite{han_ee_service_placement_mec_2022} formulates a joint optimization problem of service placement and Small cell Base Station power. The authors aim for a flexible service placement in \ac{MEC} while saving energy in the base station.

Chen et al.~\cite{green_communication_chen_21} introduce a new framework for the coexistence of unlicensed heterogeneous networks (NR-U and WiFi) utilizing network slicing. Its goal is to implement an efficient network selection and resource allocation strategy to minimize energy consumption while meeting specific user demands for data rate and delay. The authors recommend an online Lyapunov optimization algorithm for optimal scheduling, which is praised for its rapid convergence and low complexity. The approach considers two timescales to mitigate the significant costs associated with energy consumption and performance decline due to frequent network switching. The paper acknowledges a trade-off between energy consumption and traffic delay, which can be managed through adjustable control parameters. For example, the proposed strategy can accommodate a delay-sensitive user by increasing downlink energy consumption to ensure timely transmission, while conserving energy for a delay-tolerant user by extending the transmission duration.

The study \cite{mesodiakaki20215g} presents a framework for managing network resources to improve the performance of the \ac{E2E} network through the allocation and routing of the \ac{VNF} with control of delays and capacity restrictions. 
The study proposes a heuristic that considers traffic from the request point to the user, maximizing the energy efficiency of the virtualized network, calculating the shortest paths, and checking whether this path satisfies the conditional rules. 
The model considers sound provisioning and heterogeneous slice configurations, including different components of the core radio and transport networks. 
To allocate resources across network slices, the \ac{NSMF} is used in the instantiation phase supported by the Placement Service, and through algorithms, it establishes the resources that will be allocated, calculating optimization and managing the slice life cycle. 

In Zhou \cite{zhou2020dynamic}, a dynamic network slice energy-efficient deployment approach is presented. The solution uses a traffic prediction algorithm for network slices coupled with a \acf{VNF} placement and scaling strategy to deploy slices proactively. The strategy of the proposed method is based on resource sharing considering \ac{VNF} placement and \ac{VNF} traffic prediction. Authors argue that their method (NSD-CECM) reduces energy consumption compared to other methods (HSR-RSV, ESFC, and VPCM).

The document~\cite{aba_ns_orchestrator_2024} presents an enhanced scheduling approach for network slicing in 5G, leveraging \ac{K8s} and a Genetic Algorithm-based strategy to improve \ac{VNE}. 
The authors identify inefficiencies in \ac{K8s}' default scheduler, which deploys slices pod by pod, often leading to wasted resources (local optima) and increased energy consumption due to incomplete deployments. 
Instead, their proposed scheduler evaluates the feasibility of deploying an entire slice before committing resources, thereby improving slice acceptance ratio, deployment time, and energy efficiency. 
This strategy directly contributes to Path/Route Optimization in network slicing by optimizing slice placement and resource utilization, ensuring efficient orchestration within the network infrastructure.

The study \cite{chen2022designing} presents a scheduling mechanism for a \ac{VNF} orchestration system based on Kubernetes, as a solution for the efficient use of energy in network slices. The proposed solution uses a reinforcement learning algorithm (Q-Learning) to make decisions about the positioning of \acp{VNF} in the network slices, to achieve energy savings.

The document~\cite{blanco_luis_orchestration_2024} proposes a distributed \ac{AI}/\ac{ML}-based network management framework aligned with the MonB5G project. 
By leveraging \ac{FL}, the framework decentralizes management and orchestration functions to improve scalability and sustainability in 6G networks. 
The experimental results demonstrate significant energy savings (over 10x compared to centralized approaches) through proactive resource allocation and prediction mechanisms, specifically CPU scaling for \ac{VR} streaming servers.
Moreover, the paper addresses slice orchestration and resource management across domains (RAN, Core, and Edge) to minimize energy consumption.
The proposed solution optimizes paths and routes of network traffic while ensuring service quality, particularly in resource-intensive applications like \ac{VR} video streaming.

\subsection{Slice Operation Optimization}
\label{subsec:opt_level_usage}

\acf{EE} can be achieved in \ac{NS} also by optimizing the slice usage. 
Works we found in the literature are listed in Table~\ref{tab:usage_optimization}. 
In this sense, the idea is to reduce resource wastage (e.g., idle devices, overallocation). 
Methods that apply this strategy lean on the scope of the slice itself. 
Rather than optimizing the overall resource distribution and management, such methods optimize already granted resources (e.g., caching, load balancing, usage prediction, sleep mode).
Therefore, methods allocating resources for an optimized route, for example, belong to the \textit{Path/Route Optimization level} (Section~\ref{subsec:opt_level_orchestration}).

\begin{table*}[!ht]
\centering
\caption{Publications achieving \ac{EE} on the slice usage optimization level.} \label{tab:usage_optimization}
\renewcommand{\arraystretch}{1.3} 
\setlength{\arrayrulewidth}{0.8pt} 
\rowcolors{2}{gray!20}{white} 
\begin{tabular}{|
    >{\centering\arraybackslash}m{1cm} |
    >{\centering\arraybackslash}m{2cm} |
    >{\centering\arraybackslash}m{2cm} |
    >{\centering\arraybackslash}m{2cm} |
    >{\centering\arraybackslash}m{4cm} |
    >{\centering\arraybackslash}m{4cm} |}
    \hline
    \textbf{Ref.} & \textbf{Network Segment} & \textbf{3GPP Phase} & \textbf{AI Method} & \textbf{Other Methods} & \textbf{Technique} \\
    \hline

    \cite{thantharate_eco6g_2022} & RAN & Operation & \acs{DNN}, \acs{TL} & \acs{ARIMA}, \acs{ETS} & Traffic Load Prediction \& Sleep Mode \\
    \cite{yaser_2022} & RAN & Commissioning & \acs{DRL} & --- & Resource Allocation \\
    \cite{xing_joint_2022} & H-CRAN & Commissioning & \acs{DRL} & Greedy algorithm & Resource Allocation \\
    \cite{miao_ddqn_based_2022} & RAN (Sidelink Edge) & Operation & \acs{DDQN} & --- & Resource Allocation \\
    \cite{jayanthi_user_association_2021} & RAN, Edge & Operation & --- & Genetic algorithm & User Association Control \\
    \cite{ee_communication_tun_20} & \acs{MEC} & Operation & --- & \acs{BCD} & Task Offloading \\
    \cite{Bolla_FITCE_2022} & E2E & Operation & --- & Dataset analysis & Workload Control \\
    \cite{Laroui_IWCMC_2020} & Edge & Operation & \acs{RL}, \acs{DRL} & --- & Resource Allocation \& Elasticity \\
    \cite{Ravindran_EESO_2020} & RAN & Operation & --- & \acs{EESO} algorithm & Resource Allocation \\
    \cite{zhang2023energy} & F-RAN & Operation & --- & Problem decomposition & Caching \& Resource Allocation \\
    \cite{tran_duc_dung_2024} & RAN, Core & Operation & Multi-Agent DDQN & Joint Optimization Framework & Resource Allocation and usage \\
    \hline
\end{tabular}
\end{table*}

Thantharate et al.\  \cite{thantharate_eco6g_2022} propose a data-driven \ac{DNN} method with \ac{TL}, \ac{ARIMA} and \ac{ETS} to reduce energy consumption in 5G and \ac{B5G} networks. 
The approach forecasts traffic load and uses the estimated load to evaluate energy efficiency and OPEX savings. 
The authors claim that the proposed ECO6G method presents advantages concerning other OPEX saving methods. 
In this sense, energy efficiency is achieved through a usage strategy, enabling network management (e.g., putting base stations in sleep mode) according to usage prediction.

The authors in \cite{yaser_2022} combine \ac{DL} and \ac{DRL} for decision-making on resource allocation on a large and small time scale, respectively. 
Different than most reviewed studies, this work considers not only the radio resources but also the power resources to rate-based and resource-based users in the \ac{RAN}. 
The proposed method uses traffic patterns to predict power and \acp{RB} allocation for each user in the slice. 
Simulation results show better energy efficiency and overall performance than other methods published in the literature.

Also concerned with radio resource allocation, the authors in \cite{xing_joint_2022} approach a multi-objective problem by maintaining \ac{QoS} and \ac{EE} in a \acf{H-CRAN} environment. 
The authors claim to achieve a better \ac{QoS} provisioning for the slices while preserving energy consumption compared to other baseband solutions.

Miao et al.\  in \cite{miao_ddqn_based_2022} propose a vehicular network (V2X) sideline resource allocation (outside the signal coverage of the 5G base station) method that maximizes energy efficiency and meets communications latency constraints using the \ac{DRL} \ac{DDQN} algorithm. The communication resources area is allocated using the dedicated V2V spectrum.

The document~\cite{tran_duc_dung_2024} addresses energy efficiency challenges in 5G networks supporting heterogeneous services, such as \ac{eMBB}, \ac{mMTC}, and \ac{URLLC}. 
It proposes a joint optimization framework for subchannel assignment and power allocation using a cooperative multi-agent deep reinforcement learning (MA-DDQN) approach. 
By dynamically optimizing resource allocation, the approach ensures energy-efficient operation of \ac{H-NOMA} networks while maintaining \ac{QoS} requirements. 
The authors achieve energy efficiency optimization based on a strategy leveraging real-time adjustments and advanced machine learning techniques. 
Their method dynamically adjusts operations during a deployed slice's life cycle, improving the resource usage during the slice's operation.

Laroui et al.\ \cite{Laroui_IWCMC_2020} introduce novel resource allocation algorithms for \acp{VNF} in edge computing environments. These algorithms, based on \ac{ILP}, \ac{RL}, and \ac{DRL}, focus on optimizing server utilization, placement time, and energy consumption. The paper showcases how these algorithms can efficiently handle different network scales and complexities, demonstrating their scalability and cost-effectiveness in terms of energy use and server resource optimization. This work is particularly relevant in the context of 5G networks and the increasing demand for efficient resource allocation in edge computing scenarios.

The authors in \cite{zhang2023energy} considered dynamic scenarios for \acp{F-RAN} where the optimization is performed for each time slot, achieving a flexible control both on \ac{EE} and delay performance by adjusting the radio, caching, computing, and virtualization parameters. 
They propose a low-complexity \ac{EE} optimization algorithm for virtual resource allocation, by decomposing into two sub-problems, the allocation of caching and computing resources and the sub-carrier assignment and power allocation. 
They also adopt a stochastic model to accommodate virtual resources, time-varying channel conditions and random data arrivals in the simulation analysis.
Their simulation results show a 30\% improvement in \ac{EE} under the condition that the guaranteed delay threshold is two slots (i.e., 1 ms).

The study \cite{jayanthi_user_association_2021} investigates power consumption and energy efficiency in a multi-tenancy scenario for NS in \acp{HetNet}.
The user association problem presumes that the network slices were already created and are currently operational. 
Thus, the authors use a genetic algorithm to allocate resources for admitted users outperforming baseline schemes in terms of fairness, \ac{QoS}, and power consumption.

In the document~\cite{ee_communication_tun_20}, the authors investigate how to offload computation tasks to the \ac{MEC} in an energy-efficient manner. By adopting \ac{MEC}, mobile devices' energy and computation capacity is improved. Therefore, the authors formulate and solve the problem using the \ac{BCD} algorithm. The authors provide numerical results to compare the benefits of their approach over classical schemes.

Bolla et al.\ \cite{Bolla_FITCE_2022} propose an AI framework for 6G networks to enhance energy efficiency. This framework aims to create a tight relationship between the workload generated by applications in network slices and the energy consumed by the infrastructure. It introduces a high-level architecture and algorithmic roles for effective implementation. The authors argue that this method can result in considerable energy conservation while satisfying application demands, contributing to sustainable 6G networks.

Ravindran et al.\  \cite{Ravindran_EESO_2020} propose an algorithm called \acf{EESO} to achieve energy efficiency by optimizing the resources required/allocated by user services per sub-slice in 5G network. The \ac{EESO} algorithm is responsible for the optimal allocation of system resources for the services of users connected in multi-sub-slice and the energy efficiency for the services assigned to a user connected by multi-sub-slice.

\section{Taxonomy Discussion \& Directions}
\label{sec:trends_challenges}

This section presents key discussions and reflections that emerged during the development and evaluation of the proposed taxonomy for energy efficiency in network slicing.
Grounded in the initial research question, \textit{``How can one contribute to increasing energy efficiency in network slicing?''}, this discussion critically analyzes methodological choices, identifies practical and theoretical challenges, and explores the broader implications of the results.
By examining the limitations of current approaches, the complexity of implementation in real-world scenarios, and the potential of advanced technologies such as AI-driven digital twins, this section aims to deepen the reader’s understanding of the taxonomy’s contributions and to stimulate future research.
Additionally, it broadens the debate to encompass sustainability considerations, highlighting energy-efficient network slicing as a promising yet underexplored path toward environmentally and socially responsible communication infrastructures.

In this work, we set efforts on a comprehensive classification, introducing a clear and structured taxonomy for energy efficiency in network slicing, which facilitates a more in-depth understanding of existing approaches and guides future research. 
From our review of the literature and the review of methods and approaches, we understand that current techniques suit one of the three pyramid levels (Figure~\ref{fig:pyramid_taxonomy}): \textit{Infrastructure Optimization}, \textit{Path/Route Optimization}, and \textit{Slice Usage Optimization}.
The literature seems to be concerned about the RAN at the infrastructure level, especially with radio spectrum efficiency. 
Furthermore, F-RAN, \ac{MEC}, and different functional splits have been well investigated in terms of resource arrangement and allocation.
Moreover, none of the papers we found in our review mention infrastructure-level optimizations at the \acf{CN} segment. 
Since this segment is mainly responsible for virtualized network functions, energy efficiency at the infrastructure level often falls back to cloud computing or data center infrastructures.

\textit{Resource allocation} is a recurring theme in \ac{EE} research related to network slicing.
However, we observed that the term encompasses diverse techniques, depending on the type of resource being optimized (for example, function chaining versus \ac{RAN} or edge resources).
This ambiguity extends to other topics as well, such as usage prediction and user admission control.
Thus, previous classifications based solely on thematic categories fail to reveal the actual mechanisms through which \ac{EE} is achieved.
Our taxonomy addresses this shortcoming by explicitly exposing the strategies used to improve energy efficiency in \ac{NS}.
We hope this structured approach can serve as both a reference and a foundation for future research in this field.

While the proposed taxonomy provides a structured understanding of energy-efficient strategies in \acf{NS}, it is crucial to acknowledge the practical challenges that network operators face when attempting to implement these solutions in real-world scenarios.
First, the integration of \ac{EE} strategies often requires significant architectural changes, especially at the \textit{Infrastructure Optimization Level}, where methods such as edge server deployment, green hardware adoption, or cell densification demand both financial investment and logistical planning.
Furthermore, many techniques assume a high degree of control and observability over the network, which may not be feasible for operators constrained by legacy systems or vendor-specific hardware limitations.

At the \textit{Path/Route Optimization Level}, the orchestration mechanisms that enable energy-aware resource sharing and traffic management rely heavily on accurate, real-time monitoring and centralized control.
This introduces concerns regarding scalability, latency, and the robustness of inter-slice coordination.
In practice, implementing such orchestration requires standardized interfaces between the slicing controller, management plane, and the data plane (many of which are still evolving within 3GPP and ETSI specifications).

The \textit{Slice Operation Optimization Level} also presents operational challenges, particularly related to workload forecasting and dynamic scaling.
Techniques based on machine learning or workload analysis require large datasets, extensive tuning, and can introduce computational overhead that partially offsets energy savings.
Moreover, dynamically managing sleep modes or resource throttling demands fine-grained control at the hypervisor, container, or \ac{VNF} level (capabilities that may not be uniformly supported across commercial platforms).

Lastly, the lack of unified evaluation metrics and benchmarking frameworks for energy efficiency in \ac{NS} makes it difficult for operators to assess the cost-benefit ratio of deploying such strategies.
Addressing these implementation barriers will require collaboration between industry stakeholders, standardization bodies, and the research community to bridge the gap between theoretical advancements and practical deployment.

It is worth mentioning that we did not find any technique (or approach combining multiple techniques) that suits all three levels of the pyramid simultaneously. 
In 2016, the authors in~\cite{buzzi_survey_ee_techniques_5G_2016} pointed out the need for a holistic approach that combined multiple energy-efficient techniques. 
However, little has changed since then. Our understanding is that the community is concerned about very specific challenges. 
In this sense, we believe that combining multiple techniques and technologies from each of the pyramid levels is an important step in understanding their final impact and overall benefits.

Developing such a holistic approach would require the creation of cross-layer optimization frameworks that dynamically adjust the placement of network functions, routing policies, and real-time resource allocation based on energy constraints. 
To achieve this, recent studies investigate the usage of \ac{AI}-driven digital twins towards 6G networks~\cite{dig_twin_2024, dig_twin_6g_2024, dig_twin_5g_adm_control_2024}.
In this sense, a digital twin can be understood as a virtual replica of the physical network continuously used to monitor, predict, and optimize resource allocation and energy consumption based on real-time traffic demand and environmental conditions. 
\ac{DRL} is often integrated with digital twins to adapt slicing decisions, with the aim of optimizing resource usage and maintaining \ac{QoS} constraints.
However, integrating digital twins into energy-efficient network slicing remains an open challenge, particularly regarding scalability, interoperability, and real-time decision-making. 
Future research should focus on developing efficient \ac{AI} models that can predict and optimize network energy consumption with low computational overhead.

Another observation worth highlighting is that current energy-efficient slicing approaches often assume a centralized control plane managed by a single network operator.
However, real-world deployments will involve federated energy-aware resource allocation, requiring coordination between multiple infrastructure providers, such as \acp{MNO}, \acp{MVNO}, and private networks from different geographical regions.
Therefore, multi-tenancy and multi-domain energy efficiency are still challenges, not only technical but also from a standardization perspective.

Finally, the energy efficiency research topic is directly related to sustainability contributions. 
In this regard, to the best of our knowledge, the current state-of-the-art lacks a comprehensive study on the contributions and overall impacts that \acf{NS} may bring to communication networks.
We recognize that the ongoing evolution of these networks can have profound implications for human society.
Thus, in this discussion, we aim to go beyond purely technical aspects of energy-saving mechanisms.
We seek to emphasize the importance of long-term sustainability, highlighting that this topic remains underexplored and offers ample opportunities for impactful research.
In this regard, future studies could examine not only the direct contributions of \ac{NS}, but also, more broadly, the sustainability implications of advancements in communication infrastructures.
To contribute to this agenda, we outline several potential contributions of \ac{NS}, particularly in relation to the environmental, economic, and social pillars of sustainability.
These contributions, along with the mechanisms and strategies identified in this work, are summarized in Table~\ref{tab:ns_sustainability}.

\begin{table*}[htb]
\centering
\caption{\ac{NS} and its prospective influence in sustainability pillars.} \label{tab:ns_sustainability}
\renewcommand{\arraystretch}{1.3}
\begin{tabular}{@{}>{\color{black}}p{3cm} >{\color{black}}p{6.5cm} >{\color{black}}p{5.5cm}@{}}
    \toprule
    \textbf{Sustainability Pillar} & \textbf{\ac{NS} Contribution} & \textbf{Mechanism/Strategy} \\
    \midrule

    \multirow{2}{*}{\parbox[c][3cm][c]{3cm}{\centering\textbf{Environmental}}}
    & Reduces overall energy consumption and emissions 
    & \begin{itemize}
        \item Dynamic VNF placement and scaling 
        \item Energy-aware orchestration (e.g., sleep modes)
        \item Efficient RAN usage
      \end{itemize} \\
    \cline{2-3}
    & Supports integration with renewable-powered infrastructure  
    & \begin{itemize}
        \item Edge/cloud nodes powered by solar/wind energy
        \item Load-aware function scheduling
      \end{itemize} \\
    \midrule

    \multirow{2}{*}{\parbox[c][2.5cm][c]{3cm}{\centering\textbf{Economic}}}
    & Lowers \ac{OPEX} through infrastructure sharing 
    & \begin{itemize}
        \item Multi-tenancy
        \item \ac{NSaaS}
        \item On-demand resource allocation
      \end{itemize} \\
    \cline{2-3}
    & Enables new business models for vertical industries 
    & \begin{itemize}
        \item SLA-driven services
        \item Custom slices for \ac{IoT}, health, etc.
      \end{itemize} \\
    \midrule

    \multirow{2}{*}{\parbox[c][1.8cm][c]{3cm}{\centering\textbf{Social}}}
    & Improves access to digital services in underserved areas 
    & \begin{itemize}
        \item Rural \ac{FWA} using low-cost NS solutions
      \end{itemize} \\
    \cline{2-3}
    & Enhances quality and availability of public services 
    & \begin{itemize}
        \item Dedicated slices for healthcare, education, public safety
      \end{itemize} \\
    \bottomrule
\end{tabular}
\end{table*}

\section{Final Considerations}
\label{sec:conclusion}

This survey considers how we can contribute to addressing the increase of energy efficiency in network slicing, with the goal of reducing ICT's carbon footprint.

The contributions presented initially include the definition of a new approach-driven taxonomy for energy-efficient network slicing. This taxonomy contributes to a more general understanding of what is being done by the specific energy-efficient solutions provided by the research community. Instead of focusing on the end algorithm or method used, the taxonomy places the proposed solutions  for energy-efficient network slicing on different optimization levels that reflect the main approach used (infrastructure, topology, or slice). Concomitantly, this survey analyzes the related papers and maps them on the proposed taxonomy.

One relevant trend identified by our taxonomic classification is that energy efficiency in network slicing is gradually becoming a systemic approach. We mean that existing solutions tend to consider multiple optimization-level methods, ranging from the decision to place resources in the infrastructure to optimizing resource usage at the slice optimization level.

This suggests that a more holistic approach is the best way to save energy and reduce carbon footprint. This trend also suggests that energy-efficient systems will involve various subsystems, and machine-learning approaches will have to be integrated learning from distributed sources of knowledge. This is indeed a new challenge to be addressed by the research community. 

\section*{List of Acronyms}



\begin{acronym}[XGBoost]
\acro{3GPP}{3rd Generation Partnership Project}
\acro{6G}{sixth-generation}
\acro{5G}{fifth-generation}
\acro{A2C}{Advantage Actor-Critic}
\acro{ACM}{Association for Computing Machinery}
\acro{AN}[AN]{Access Network}
\acro{AGNS}{Automatic Generation of Network Slices}
\acro{AI}{Artificial Intelligence}
\acro{AP}{Access Point}
\acro{ARIMA}{Autoregressive Integrated Moving Average}
\acro{B2C}[B2C]{Business to Consumer}
\acro{B2B}[B2B]{Business to Business}
\acro{B5G}{Beyond 5G}
\acro{BCD}{block coordinate descent}
\acro{BE}{Best Effort}
\acro{BS}{Base Station}
\acro{C-RAN}{Cloud-RAN}
\acro{CBDA}{Chain-Based Data Aggregation}
\acro{CC}{Cognitive Cycles}
\acro{CN}{Core Network}
\acro{CNN}{Convolutional Neural Network}
\acro{CPU}{Central Processing Unit}
\acro{CSI}{Channel State Information}
\acro{CT}{Communication Technology}
\acro{CU}{Central Unit}
\acro{DCIE}{Data Center Infrastructure Efficiency}
\acro{DCMAB}{Deep Contextual MAB}
\acro{DDPG}{Deep Deterministic Policy Gradient}
\acro{DDQN}{Double Deep Q-Network}
\acro{DQL}{Deep Q-Learning}
\acro{DQN}{Deep Q-Network}
\acro{DL}{Deep Learning}
\acro{DLNN}{Deep Learning Neural Network}
\acro{DRL}{Deep Reinforcement Learning}
\acro{DNN}{Deep Neural Network} 
\acro{DU}{Distributed Unit}
\acro{DPI}{Deep Packet Inspection}
\acro{E2E}{end-to-end}
\acro{EC}{Edge Controller}
\acro{EE}{Energy Efficiency}
\acro{EESO}{Energy Efficient System-resource Optimization}
\acro{eMBB}{enhanced Mobile Broadband}
\acro{EPDA}{Exponential Power Descent Algorithm}
\acro{ETS}{Exponential Smoothing}
\acro{ETSI}{European Telecommunications Standards Institute}
\acro{FL}{Federated Learning}
\acro{F-AP}{Fog Radio Access Point}
\acro{F-RAN}{Fog-RAN}
\acro{F-UE}{Fog-UE}
\acro{FWA}{Fixed Wireless Access}
\acro{GA}{Genetic Algorithm}
\acro{GBR}{Gradient Boosting Regressor}
\acro{GCN}{Graph Convolutional Network}
\acro{GHG}{Greenhouse Gas}
\acro{H2H}{Human-to-Human}
\acro{H-CRAN}{Heterogeneous Cloud Radio Access Network}
\acro{HDBSCAN}{Hierarchical Density Based Spatial Clustering}
\acro{HetNet}{Heterogeneous Network}
\acro{H-NOMA}{heterogeneous non-orthogonal multiple access}
\acro{HW}{Holt-Winters}
\acro{ICT}{Information and Communication Technology}
\acro{IEEE}{Institute of Electrical and Electronics Engineers}
\acro{IETF}{Internet Engineering Task Force}
\acro{ILP}{Integer Linear Programming}
\acro{IP}{Internet Protocol}
\acro{InP}{Infrastructure Provider}
\acro{IoT}{Internet of Things}
\acro{IIoT}{Industrial Internet of Things}
\acro{ISP}{Internet Service Provider}
\acro{ITU-T}{International Telecommunication Union Telecommunication Standardization Sector}
\acro{K8s}{Kubernetes}
\acro{KKT}{Karush–Kuhn–Tucker}
\acro{KPI}{Key Performance Indicator}
\acro{LCM}{Life Cycle Management}
\acro{LIME}{Local Interpretable Model-Agnostic Explanations}
\acro{LR}{Logistic Regression}
\acro{LSTM}{Long Short Term Memory}
\acro{LTE}{Long Term Evolution}
\acro{M2M}{Machine-to-Machine}
\acro{MAB}{Multi-Armed Bandit}
\acro{MANO}{Management and Orchestration}
\acro{MBLP}{Mixed Binary Linear Program}
\acro{MEC}{Multi-access Edge Computing}
\acro{MDP}{Markov Decision Process}
\acro{MILFP}{mixed-integer linear fractional optimization problem}
\acro{mIoT}{massive IoT}
\acro{MILP}{mixed-integer linear programming}
\acro{MINLP}{mixed-integer non-linear programming}
\acro{mMTC}{Machine Type Communication}
\acro{MNO}{Mobile Network Operator}
\acro{MPLS}{Multiprotocol Label Switching}
\acro{MVNO}{Mobile Virtual Network Operator}
\acro{ML}{Machine Learning}
\acro{MVNO}{Mobile Virtual Network Operator}
\acro{N3AC}{Neural Network Admission Control}
\acro{NECOS}{Novel Enablers for Cloud Slicing}
\acro{NFV}{Network Function Virtualization}
\acro{NGMN}{Next-Generation Mobile Networks}
\acro{NLP}{Natural Language Processing}
\acro{NN}{Neural Network}
\acro{NOMA}{Non-Orthogonal Multiple Access}
\acro{NR}{New Radio}
\acro{NS}{Network Slicing}
\acro{NSaaS}{Network Slice-as-a-Service}
\acro{NSMF}{Network Slice Management Function}
\acro{NSS}{Network Slice Subnet}
\acro{NS-3}{Network Simulator 3}
\acro{NSP}{Network Service Provider} 
\acro{OMA}{Orthogonal Multiple Access}
\acro{ONETS}{Online NETwork Slice Broker}
\acro{O-RAN}{Open RAN}
\acro{OPEX}{Operating expenses}
\acro{PI}{Permutation Importance}
\acro{PRB}{Physical Resource Block}
\acro{PoP}{Point of Presence}
\acro{PT}{Prospect Theory}
\acro{PUE}{Power Usage Effectiveness}
\acro{QCI}{QoS Class Identifier}
\acro{QL}{Q-Learning}
\acro{QoE}{Quality of Experience}
\acro{QoT}{Quality of Transmission}
\acro{QoS}{Quality of Service}
\acro{RIC}{RAN Intelligent Controller}
\acro{SFC}{service function chain}
\acro{UAV}{Unmanned Aerial Vehicle}
\acro{UE}{User Equipment}
\acro{URLLC}{Ultra-Reliable and Low-Latency Communications}
\acro{RAM}{Random Access Memory}
\acro{RAN}{Radio Access Network}
\acro{RAT}{Radio Access Technology}
\acro{RB}{Resource Block}
\acro{RF}{Random Forest}
\acro{RFR}{Random Forest Regressor}
\acro{RL}{Reinforcement Learning}
\acro{RMEC}{Radio Multi-access Edge Computing}
\acro{RU}{Radio Unit}
\acro{SACRA}{Slicing Aware Clustering and Resource Allocation}
\acro{SARSA}{State-Action-Reward-State-Action}
\acro{SCA}{Successive Convex Approximation}
\acro{SDN}{Software-Defined Network}
\acro{SFI2}{Slicing Future Internet Infrastructures}
\acro{SLA}{Service Level Agreement}
\acro{SLAW}{Self-similar least-action human walk}
\acro{SDO}{Standards Developing Organization}
\acro{SMDP}{Semi-Markov Decision Process}
\acro{SON}{Self-Organized Network}
\acro{SHAP}{SHapely Additive Explanations}
\acro{SL}{Supervised Learning}
\acro{SMO}{Service Management and Orchestration}
\acro{SP}{Slice Provider}
\acro{SVM}{Support Vector Machine}
\acro{SBLR}{Sparse Bayesian Linear Regression}
\acro{TL}{Transfer Learning}
\acro{TN}{Transport Network}
\acro{UCB}{Upper Confidence Bound}  
\acro{UE}[UE]{User Equipment}
\acro{UL}{Unsupervised Learning}
\acro{V2I}{Vehicle-to-Infrastructure}
\acro{V2N}{Vehicle-to-Network}
\acro{V2P}{Vehicle-to-Pedestrian}
\acro{V2V}{Vehicle-to-Vehicle}
\acro{V2X}{Vehicle-to-Everything}
\acro{vBS}{virtual Base Station}
\acro{VM}{Virtual Machine}
\acro{VNE}{Virtual Network Embedding}
\acro{VNF}[VNF]{Virtualized Network Function}
\acro{VNFD}[VNFD]{Virtualized Network Function Descriptor}
\acro{VR}{Virtual Reality}
\acro{vRAN}{Virtual RAN}
\acro{VS}{Virtual Slicing}
\acro{WET}{Wireless Power Transmission}
\acro{Wi-Fi}{Wireless Fidelity}
\acro{XAI}[XAI]{eXplainable Artificial Intelligence}
\acro{XGBoost}{Extreme Gradient Boosting}
\acro{ZSM}{Zero touch network \& Service Management}

\end{acronym}

\bibliographystyle{IEEEtran}
\begingroup
\sloppy
\bibliography{ref.bib}

\begin{thebibliography}{10}
\providecommand{\url}[1]{#1}
\csname url@samestyle\endcsname
\providecommand{\newblock}{\relax}
\providecommand{\bibinfo}[2]{#2}
\providecommand{\BIBentrySTDinterwordspacing}{\spaceskip=0pt\relax}
\providecommand{\BIBentryALTinterwordstretchfactor}{4}
\providecommand{\BIBentryALTinterwordspacing}{\spaceskip=\fontdimen2\font plus
\BIBentryALTinterwordstretchfactor\fontdimen3\font minus \fontdimen4\font\relax}
\providecommand{\BIBforeignlanguage}[2]{{%
\expandafter\ifx\csname l@#1\endcsname\relax
\typeout{** WARNING: IEEEtran.bst: No hyphenation pattern has been}%
\typeout{** loaded for the language `#1'. Using the pattern for}%
\typeout{** the default language instead.}%
\else
\language=\csname l@#1\endcsname
\fi
#2}}
\providecommand{\BIBdecl}{\relax}
\BIBdecl

\bibitem{freitag_2021}
C.~Freitag, M.~Berners-Lee, K.~Widdicks, B.~Knowles, G.~S. Blair, and A.~Friday, ``The real climate and transformative impact of {ICT}: A critique of estimates, trends, and regulations,'' \emph{Patterns}, vol.~2, no.~9, p. 100340, 2021.

\bibitem{huawei_andrae_update_2020}
A.~Andrae, ``New perspectives on internet electricity use in 2030,'' \emph{Engineering and Applied Science Letters}, vol.~3, pp. 19--31, 06 2020.

\bibitem{huawei_andrae_2015}
\BIBentryALTinterwordspacing
A.~S.~G. Andrae and T.~Edler, ``On global electricity usage of communication technology: Trends to 2030,'' \emph{Challenges}, vol.~6, no.~1, pp. 117--157, 2015. [Online]. Available: \url{https://www.mdpi.com/2078-1547/6/1/117}
\BIBentrySTDinterwordspacing

\bibitem{lorincz_greener_2019}
J.~Lorincz, A.~Capone, and J.~Wu, ``\BIBforeignlanguage{en}{Greener, {Energy}-{Efficient} and {Sustainable} {Networks}: {State}-{Of}-{The}-{Art} and {New} {Trends}},'' \emph{\BIBforeignlanguage{en}{Sensors}}, vol.~19, no.~22, p. 4864, Jan. 2019.

\bibitem{martins_enhancing_2023}
J.~S.~B. Martins, T.~C. Carvalho, R.~Moreira, C.~B. Both, A.~Donatti, J.~H. Corrêa, J.~A. Suruagy, S.~L. Corrêa, A.~J.~G. Abelem, M.~R.~N. Ribeiro, J.-m.~S. Nogueira, L.~C.~S. Magalhães, J.~Wickboldt, T.~C. Ferreto, R.~Mello, R.~Pasquini, M.~Schwarz, L.~N. Sampaio, D.~F. Macedo, J.~F. De~Rezende, K.~V. Cardoso, and F.~De~Oliveira~Silva, ``Enhancing {Network} {Slicing} {Architectures} {With} {Machine} {Learning}, {Security}, {Sustainability} and {Experimental} {Networks} {Integration},'' \emph{IEEE Access}, vol.~11, pp. 69\,144--69\,163, 2023.

\bibitem{alamu_survey_techniques_hetnets_2020}
\BIBentryALTinterwordspacing
O.~Alamu, A.~Gbenga-Ilori, M.~Adelabu, A.~Imoize, and O.~Ladipo, ``Energy efficiency techniques in ultra-dense wireless heterogeneous networks: An overview and outlook,'' \emph{Engineering Science and Technology, an International Journal}, vol.~23, no.~6, pp. 1308--1326, 2020. [Online]. Available: \url{https://www.sciencedirect.com/science/article/pii/S2215098619328745}
\BIBentrySTDinterwordspacing

\bibitem{buzzi_survey_ee_techniques_5G_2016}
S.~Buzzi, C.-L. I, T.~E. Klein, H.~V. Poor, C.~Yang, and A.~Zappone, ``A survey of energy-efficient techniques for 5g networks and challenges ahead,'' \emph{IEEE Journal on Selected Areas in Communications}, vol.~34, no.~4, pp. 697--709, 2016.

\bibitem{larsen_toward_2023}
L.~M.~P. Larsen, H.~L. Christiansen, S.~Ruepp, and M.~S. Berger, ``Toward {Greener} {5G} and {Beyond} {Radio} {Access} {Networks}—{A} {Survey},'' \emph{IEEE Open Journal of the Communications Society}, vol.~4, pp. 768--797, 2023.

\bibitem{setiawan_energy-efficient_2024}
I.~Setiawan, B.~Kar, and S.-H. Shen, ``Energy-{Efficient} {Softwarized} {Networks}: {A} {Survey},'' \emph{Transactions on Network and Service Management}, pp. 1--21, 2024.

\bibitem{5G-NS-NGMN}
\BIBentryALTinterwordspacing
NGMN, ``{5G} white paper, version 1.0,'' NGMN Alliance, Tech. Rep., Feb. 2015. [Online]. Available: \url{https://ngmn.org/wp-content/uploads/NGMN_5G_White_Paper_V1_0.pdf}
\BIBentrySTDinterwordspacing

\bibitem{5G-NS-VTC}
M.~Iwamura, ``{NGMN} view on {5G} architecture,'' in \emph{2015 IEEE 81st Vehicular Technology Conference (VTC Spring)}, 2015, pp. 1--5.

\bibitem{3gpp_2023_ts28530_18_0}
{3GPP}, ``{Management and orchestration; Concepts, use cases and requirements ({Release} 18)},'' Available online: \url{https://www.3gpp.org/ftp/Specs/archive/28_series/28.530/}, 2023, accessed on 12 March 2024.

\bibitem{slice_how_to_2019}
W.~Lee, T.~Na, and J.~Kim, ``How to create a network slice? - a {5G} core network perspective,'' in \emph{2019 21st International Conference on Advanced Communication Technology (ICACT)}, 2019, pp. 616--619.

\bibitem{nso_survey}
\BIBentryALTinterwordspacing
N.~F. {Saraiva de Sousa}, D.~A. {Lachos Perez}, R.~V. Rosa, M.~A. Santos, and C.~{Esteve Rothenberg}, ``Network service orchestration: A survey,'' \emph{Computer Communications}, vol. 142-143, pp. 69--94, 2019. [Online]. Available: \url{https://www.sciencedirect.com/science/article/pii/S0140366418309502}
\BIBentrySTDinterwordspacing

\bibitem{5g_ppp_5g_vision}
\BIBentryALTinterwordspacing
5GPP, ``{5G} vision: the next generation of communication networks and services,'' The 5G infrastructure public private partnership, Tech. Rep., Feb. 2015. [Online]. Available: \url{http://5g-ppp.eu/wp-content/uploads/2015/02/5G-Vision-Brochure-v1.pdf}
\BIBentrySTDinterwordspacing

\bibitem{foukas_survey_ns_2017}
X.~Foukas \emph{et~al.}, ``{Network Slicing in 5G: Survey and Challenges},'' \emph{IEEE Communications Magazine}, vol.~55, no.~5, pp. 94--100, 2017.

\bibitem{donatti_survey_2023}
A.~Donatti, S.~L. Correa, J.~S.~B. Martins, A.~Abelem, C.~B. Both, F.~Silva, J.~A. Suruagy, R.~Pasquini, R.~Moreira, K.~V. Cardoso, and T.~C. Carvalho, ``Survey on {Machine} {Learning}-{Enabled} {Network} {Slicing}: {Covering} the {Entire} {Life} {Cycle},'' \emph{IEEE Transactions on Network and Service Management}, vol.~21, no.~3, pp. 1--18, 2023.

\bibitem{ml_ns_survey}
H.~P. Phyu, D.~Naboulsi, and R.~Stanica, ``Machine learning in network slicing—a survey,'' \emph{IEEE Access}, vol.~11, pp. 39\,123--39\,153, 2023.

\bibitem{shen_ai_assisted_ns_20}
X.~Shen \emph{et~al.}, ``{AI-Assisted Network-Slicing Based Next-Generation Wireless Networks},'' \emph{IEEE Open Journal of Vehicular Technology}, vol.~1, pp. 45--66, 2020.

\bibitem{moreira_intelligent_2024}
R.~Moreira, F.~d.~O. Silva, T.~C.~C. Carvalho, and J.~S.~B. Martins, ``Intelligent {Data}-{Driven} {Architectural} {Features} {Orchestration} for {Network} {Slicing},'' in \emph{Proceedings of the {International} {Workshop} on {ADVANCEs} in {ICT} {Infrastructures} and {Services}}.\hskip 1em plus 0.5em minus 0.4em\relax Hanoi, Vietnam: Université Paris-Saclay Évry, Feb. 2024, pp. 1--12.

\bibitem{chegui_complementar_2021}
H.~Chergui \emph{et~al.}, ``{Zero-Touch AI-Driven Distributed Management for Energy-Efficient 6G Massive Network Slicing},'' \emph{IEEE Network}, vol.~35, no.~6, pp. 43--49, 2021.

\bibitem{khan_2020}
H.~Khan, M.~M. Butt, S.~Samarakoon, P.~Sehier, and M.~Bennis, ``Deep learning assisted {CSI} estimation for joint {URLLC} and {eMBB} resource allocation,'' in \emph{2020 IEEE International Conference on Communications Workshops (ICC Workshops)}, 2020, pp. 1--6.

\bibitem{tang_2021}
J.~Tang, Y.~Duan, Y.~Zhou, and J.~Jin, ``Distributed slice selection-based computation offloading for intelligent vehicular networks,'' \emph{IEEE Open Journal of Vehicular Technology}, vol.~2, pp. 261--271, 2021.

\bibitem{ayala_romero_2022}
J.~A. Ayala-Romero, A.~Garcia-Saavedra, M.~Gramaglia, X.~Costa-Pérez, A.~Banchs, and J.~J. Alcaraz, ``{vrAIn}: Deep learning based orchestration for computing and radio resources in {vRANs},'' \emph{IEEE Transactions on Mobile Computing}, vol.~21, no.~7, pp. 2652--2670, 2022.

\bibitem{yu_nuo_2017}
N.~Yu, Z.~Song, H.~Du, H.~Huang, and X.~Jia, ``Multi-resource allocation in cloud radio access networks,'' in \emph{2017 IEEE International Conference on Communications (ICC)}, 2017, pp. 1--6.

\bibitem{moreira_enhancing_2023}
R.~Moreira, J.~S.~B. Martins, T.~C. M.~B. Carvalho, and F.~d.~O. Silva, ``\BIBforeignlanguage{en}{On {Enhancing} {Network} {Slicing} {Life}-{Cycle} {Through} an {AI}-{Native} {Orchestration} {Architecture}},'' in \emph{\BIBforeignlanguage{en}{Advanced {Information} {Networking} and {Applications}}}, ser. Lecture {Notes} in {Networks} and {Systems}, L.~Barolli, Ed.\hskip 1em plus 0.5em minus 0.4em\relax Cham: Springer International Publishing, 2023, pp. 124--136.

\bibitem{martinez_beltran_decentralized_2023}
E.~T. Martínez~Beltrán, M.~Q. Pérez, P.~M.~S. Sánchez, S.~L. Bernal, G.~Bovet, M.~G. Pérez, G.~M. Pérez, and A.~H. Celdrán, ``Decentralized {Federated} {Learning}: {Fundamentals}, {State} of the {Art}, {Frameworks}, {Trends}, and {Challenges},'' \emph{IEEE Communications Surveys \& Tutorials}, vol.~25, no.~4, pp. 2983--3013, 2023.

\bibitem{yaser_2022}
Y.~Azimi, S.~Yousefi, H.~Kalbkhani, and T.~Kunz, ``Energy-efficient deep reinforcement learning assisted resource allocation for {5G-RAN} slicing,'' \emph{IEEE Transactions on Vehicular Technology}, vol.~71, no.~1, pp. 856--871, 2022.

\bibitem{blanco_luis_orchestration_2024}
L.~Blanco, E.~Zeydan, S.~Barrachina-Muñoz, F.~Rezazadeh, L.~Vettori, and J.~Mangues-Bafalluy, ``A novel approach for scalable and sustainable {6G} networks,'' \emph{IEEE Open Journal of the Communications Society}, vol.~5, pp. 1673--1692, 2024.

\bibitem{blanco_chergui_2022}
H.~Chergui, L.~Blanco, and C.~Verikoukis, ``Statistical federated learning for beyond {5G SLA}-constrained {RAN} slicing,'' \emph{IEEE Transactions on Wireless Communications}, vol.~21, no.~3, pp. 2066--2076, 2022.

\bibitem{thantharate_eco6g_2022}
\BIBentryALTinterwordspacing
A.~Thantharate, A.~V. Tondwalkar, C.~Beard, and A.~Kwasinski, ``{ECO6G}: Energy and cost analysis for network slicing deployment in beyond {5G} networks,'' \emph{Sensors}, vol.~22, no.~22, 2022. [Online]. Available: \url{https://www.mdpi.com/1424-8220/22/22/8614}
\BIBentrySTDinterwordspacing

\bibitem{Bolla_FITCE_2022}
R.~Bolla, R.~Bruschi, F.~Davoli, L.~Ivaldi, C.~Lombardo, and B.~Siccardi, ``An {AI} framework for fostering {6G} towards energy efficiency,'' in \emph{2022 61st FITCE International Congress Future Telecommunications: Infrastructure and Sustainability (FITCE)}, 2022, pp. 1--6.

\bibitem{sundaramoorthy_energy_2023}
S.~Sundaramoorthy, D.~Kamath, S.~Nimbalkar, C.~Price, T.~Wenning, and J.~Cresko, ``\BIBforeignlanguage{en}{Energy {Efficiency} as a {Foundational} {Technology} {Pillar} for {Industrial} {Decarbonization}},'' \emph{\BIBforeignlanguage{en}{Sustainability}}, vol.~15, no.~12, p. 9487, Jan. 2023.

\bibitem{hafez_energy_2023}
F.~S. Hafez, B.~Sa'di, M.~Safa-Gamal, Y.~H. Taufiq-Yap, M.~Alrifaey, M.~Seyedmahmoudian, A.~Stojcevski, B.~Horan, and S.~Mekhilef, ``Energy {Efficiency} in {Sustainable} {Buildings}: {A} {Systematic} {Review} with {Taxonomy}, {Challenges}, {Motivations}, {Methodological} {Aspects}, {Recommendations}, and {Pathways} for {Future} {Research},'' \emph{Energy Strategy Reviews}, vol.~45, p. 101013, Jan. 2023.

\bibitem{patterson_what_1996}
M.~G. Patterson, ``What is {Energy} {Efficiency}?: {Concepts}, {Indicators} and {Methodological} {Issues},'' \emph{Energy Policy}, vol.~24, no.~5, pp. 377--390, May 1996.

\bibitem{anser2021energy}
Y.~Anser, J.-L. Grimault, S.~Bouzefrane, and C.~Gaber, ``Energy-aware service level agreements in {5G NFV} architecture,'' in \emph{2021 8th International Conference on Future Internet of Things and Cloud (FiCloud)}, 2021, pp. 377--382.

\bibitem{wimbadi_decarbonization_2020}
R.~W. Wimbadi and R.~Djalante, ``From {Decarbonization} to {Low} {Carbon} {Development} and {Transition}: {A} {Systematic} {Literature} {Review} of the {Conceptualization} of {Moving} {Toward} {Net}-{Zero} {Carbon} {Dioxide} {Emission} (1995–2019),'' \emph{Journal of Cleaner Production}, vol. 256, p. 120307, May 2020.

\bibitem{undesa_agenda_2030}
\BIBentryALTinterwordspacing
D.~o.~E. UNDESA United~Nations and S.~Affairs, ``\BIBforeignlanguage{en}{Sustainable development goals: The 17 {UN} goals for 2030},'' undated. [Online]. Available: \url{https://sdgs.un.org/goals}
\BIBentrySTDinterwordspacing

\bibitem{rosario_new_2023}
A.~T. Rosário and J.~C. Dias, ``\BIBforeignlanguage{en}{The {New} {Digital} {Economy} and {Sustainability}: {Challenges} and {Opportunities}},'' \emph{\BIBforeignlanguage{en}{Sustainability}}, vol.~15, no.~14, p. 10902, Jan. 2023.

\bibitem{wang_making_2021}
Z.~Wang, H.-T. Liao, J.~Lou, and Y.~Liu, ``Making {Cyberspace} {Towards} {Sustainability} {A} {Scientometric} {Review} for a {Cyberspace} that {Enables} {Green} and {Digital} {Transformation},'' in \emph{Proceedings of the 2020 {International} {Conference} on {Cyberspace} {Innovation} of {Advanced} {Technologies}}, ser. {CIAT} 2020, New York, NY, USA, Jan. 2021, pp. 394--400.

\bibitem{farhan_energy_2021}
L.~Farhan, R.~S. Hameed, A.~S. Ahmed, A.~H. Fadel, W.~Gheth, L.~Alzubaidi, M.~A. Fadhel, and M.~Al-Amidie, ``\BIBforeignlanguage{en}{Energy {Efficiency} for {Green} {Internet} of {Things} ({IoT}) {Networks}: {A} {Survey}},'' \emph{\BIBforeignlanguage{en}{Network}}, vol.~1, no.~3, pp. 279--314, Dec. 2021.

\bibitem{hu_development_2022}
J.-L. Hu, Y.-C. Chen, and Y.-P. Yang, ``\BIBforeignlanguage{en}{The {Development} and {Issues} of {Energy}-{ICT}: {A} {Review} of {Literature} with {Economic} and {Managerial} {Viewpoints}},'' \emph{\BIBforeignlanguage{en}{Energies}}, vol.~15, no.~2, p. 594, Jan. 2022, number: 2 Publisher: Multidisciplinary Digital Publishing Institute.

\bibitem{jin_energy_2012}
Y.~Jin, Y.~Wen, and Q.~Chen, ``Energy {Efficiency} and {Server} {Virtualization} in {Data} {Centers}: {An} {Empirical} {Investigation},'' in \emph{2012 {Proceedings} {IEEE} {INFOCOM} {Workshops}}, Mar. 2012, pp. 133--138.

\bibitem{loschi_energy_2015}
H.~J. Loschi, J.~Leon, Y.~Iano, E.~R. Filho, F.~D. Conte, T.~C. Lustosa, and P.~O. Freitas, ``\BIBforeignlanguage{en}{Energy {Efficiency} in {Smart} {Grid}: {A} {Prospective} {Study} on {Energy} {Management} {Systems}},'' \emph{\BIBforeignlanguage{en}{Smart Grid and Renewable Energy}}, vol.~6, no.~8, pp. 250--259, Aug. 2015.

\bibitem{xin_2017}
X.~Li, M.~Samaka, H.~A. Chan, D.~Bhamare, L.~Gupta, C.~Guo, and R.~Jain, ``Network slicing for {5G}: Challenges and opportunities,'' \emph{IEEE Internet Computing}, vol.~21, no.~5, pp. 20--27, 2017.

\bibitem{najjuuko_rural_2021}
C.~Najjuuko, G.~K. Ayebare, R.~Lukanga, E.~Mugume, and D.~Okello, ``A survey of {5G} for rural broadband connectivity,'' in \emph{2021 IST-Africa Conference (IST-Africa)}, 2021, pp. 1--10.

\bibitem{markhasin_2017}
A.~Markhasin, ``Fundamentals of the extremely green, flexible, and profitable {5G M2M} ubiquitous communications for remote e-healthcare and other social e-applications,'' in \emph{2017 International Multi-Conference on Engineering, Computer and Information Sciences (SIBIRCON)}, 2017, pp. 292--297.

\bibitem{resource_positioning_IBM_10}
Y.~Lin, L.~Shao, Z.~Zhu, Q.~Wang, and R.~K. Sabhikhi, ``Wireless network cloud: Architecture and system requirements,'' \emph{IBM Journal of Research and Development}, vol.~54, no.~1, pp. 4:1--4:12, 2010.

\bibitem{dc_max_capacity_patel_13}
J.~U. Patel, S.~J. Guercio, A.~E. Bruno, M.~D. Jones, and T.~R. Furlani, ``Implementing green technologies and practices in a high performance computing center,'' in \emph{2013 International Green Computing Conference Proceedings}, 2013, pp. 1--8.

\bibitem{dc_max_capacity_vishwakarma_15}
M.~M. Vishwakarma and M.~Manoria, ``A robust algorithm for capacity management to improve response time in virtualized data centers,'' in \emph{2015 International Conference on Computational Intelligence and Communication Networks (CICN)}, 2015, pp. 731--735.

\bibitem{energy_efficient_ethernet_22}
J.~L. Bolonhezi~Dias, L.~B. De~Almeida, and L.~C. Pessoa~Albini, ``Reducing hadoop 3.x energy consumption through energy efficient ethernet,'' in \emph{2022 IEEE International Conference on Big Data and Smart Computing (BigComp)}, 2022, pp. 9--14.

\bibitem{xiang_2020_mode_selection}
H.~Xiang \emph{et~al.}, ``Mode selection and resource allocation in sliced fog radio access networks: A reinforcement learning approach,'' \emph{IEEE Transactions on Vehicular Technology}, vol.~69, no.~4, pp. 4271--4284, 2020.

\bibitem{phyu_EE_RAN_NS_2023}
H.~P. Phyu, D.~Naboulsi, R.~Stanica, and G.~Poitau, ``Towards energy efficiency in {RAN} network slicing,'' in \emph{2023 IEEE 48th Conference on Local Computer Networks (LCN)}, 2023, pp. 1--9.

\bibitem{sen_towards_2023}
N.~Sen and A.~F. A, ``Towards {Energy} {Efficient} {Functional} {Split} and {Baseband} {Function} {Placement} for {5G} {RAN},'' in \emph{2023 {IEEE} 9th {International} {Conference} on {Network} {Softwarization} ({NetSoft})}, Jun. 2023, pp. 237--241, iSSN: 2693-9789.

\bibitem{kao_qoe_sustainability_2023}
H.-W. Kao and E.~H.-K. Wu, ``{QoE} sustainability on {5G} and beyond {5G} networks,'' \emph{IEEE Wireless Communications}, vol.~30, no.~1, pp. 118--125, 2023.

\bibitem{Hossain_NOMA_2023}
M.~A. Hossain and N.~Ansari, ``Network slicing for {NOMA}-enabled edge computing,'' \emph{IEEE Transactions on Cloud Computing}, vol.~11, no.~1, pp. 811--821, 2023.

\bibitem{hossain_numerology_2023}
A.~Hossain and N.~Ansari, ``{5G} multi-band numerology-based {TDD RAN} slicing for throughput and latency sensitive services,'' \emph{IEEE Transactions on Mobile Computing}, vol.~22, no.~3, pp. 1263--1274, 2023.

\bibitem{ismail_advances_2020}
S.~Ismail, F.~D’Andreagiovanni, H.~Lakhlef, and Y.~Imine, ``Recent advances on {5G} resource allocation problem using {PD-NOMA},'' in \emph{2020 International Symposium on Networks, Computers and Communications (ISNCC)}, 2020, pp. 1--7.

\bibitem{liu_2023}
B.~Liu, P.~Zhu, J.~Li, D.~Wang, and X.~You, ``Energy-efficient optimization in distributed massive {MIMO} systems for slicing {eMBB} and {URLLC} services,'' \emph{IEEE Transactions on Vehicular Technology}, vol.~72, no.~8, pp. 10\,473--10\,487, 2023.

\bibitem{zou2022energy}
S.~Zou, H.~Yu, W.~Wang, and W.~Ni, ``An energy-efficient framework in radio multi-access edge cloud by automatically generating network slice,'' \emph{IEEE Internet of Things Magazine}, vol.~5, no.~3, pp. 128--133, 2022.

\bibitem{ndikumana_2024}
A.~Ndikumana, K.~K. Nguyen, and M.~Cheriet, ``Renewable energy powered and {Open RAN}-based architecture for {5G} fixed wireless access provisioning in rural areas,'' \emph{IEEE Transactions on Green Communications and Networking}, vol.~8, no.~3, pp. 994--1007, 2024.

\bibitem{flexible_routing_chen_20}
W.-K. Chen, Y.-F. Liu, A.~De~Domenico, and Z.-Q. Luo, ``Network slicing for service-oriented networks with flexible routing and guaranteed e2e latency,'' in \emph{2020 IEEE 21st International Workshop on Signal Processing Advances in Wireless Communications (SPAWC)}, 2020, pp. 1--5.

\bibitem{s_joint_2023}
S.~S., S.~Mishra, and C.~Hota, ``Joint {QoS} and {Energy}-{Efficient} {Resource} {Allocation} and {Scheduling} in {5G} {Network} {Slicing},'' \emph{Computer Communications}, vol. 202, pp. 110--123, Mar. 2023.

\bibitem{mesodiakaki20215g}
A.~Mesodiakaki, M.~Gatzianas, G.~Kalfas, F.~Moscatelli, G.~Landi, N.~Ciulli, and L.~Lossi, ``{5G-COMPLETE}: End-to-end resource allocation in highly heterogeneous beyond {5G} networks,'' in \emph{2021 IEEE 4th 5G World Forum (5GWF)}, 2021, pp. 412--417.

\bibitem{ma_jinliang_slice_aware_2019}
J.~Ma, C.~Pan, C.~Yin, and X.~Li, ``Slice-aware resource management in {SDN} enabled heterogeneous cellular networks,'' in \emph{2019 IEEE/CIC International Conference on Communications in China (ICCC)}, 2019, pp. 869--874.

\bibitem{xu_2022_ee_mec}
Y.~Xu, Z.~He, Y.~Zhang, and W.~Zhou, ``Energy-efficient resource allocation for slicing-enabled multi-access edge computing,'' in \emph{2022 IEEE Smartworld, Ubiquitous Intelligence \& Computing, Scalable Computing \& Communications, Digital Twin, Privacy Computing, Metaverse, Autonomous \& Trusted Vehicles (SmartWorld/UIC/ScalCom/DigitalTwin/PriComp/Meta)}, 2022, pp. 698--705.

\bibitem{han_ee_service_placement_mec_2022}
P.~Han, Y.~Liu, X.~Zhang, and L.~Guo, ``Energy-efficient service placement based on equivalent bandwidth in cell zooming enabled mobile edge cloud networks,'' \emph{IEEE Transactions on Vehicular Technology}, vol.~71, no.~11, pp. 12\,275--12\,290, 2022.

\bibitem{oladejo_energy-efficient_2019}
S.~O. Oladejo and O.~Falowo, ``\BIBforeignlanguage{en}{An {Energy}-{Efficient} {Resource} {Allocation} {Scheme} for {5G} {Slice} {Networks}},'' in \emph{\BIBforeignlanguage{en}{Proceedings of the {Southern} {Africa} {Telecommunication} {Networks} and {Applications} {Conference} ({SATNAC})}}, South Africa, 2019, pp. 1--6.

\bibitem{green_communication_chen_21}
Q.~Chen, X.~Xu, and H.~Jiang, ``Online green communication scheduling for sliced unlicensed heterogeneous networks,'' \emph{IEEE Transactions on Vehicular Technology}, vol.~70, no.~10, pp. 10\,657--10\,670, 2021.

\bibitem{Basu_Dynamic_2022}
D.~Basu, A.~Jain, U.~Ghosh, and R.~Datta, ``{QoS}-aware dynamic network slicing and {VNF} embedding in softwarized {5G} networks,'' pp. 100--105, 2022.

\bibitem{cao_2022_RESOURCE_ALLOCATION}
H.~Cao, J.~Du, H.~Zhao, D.~X. Luo, N.~Kumar, L.~Yang, and F.~R. Yu, ``Toward tailored resource allocation of slices in {6G} networks with softwarization and virtualization,'' \emph{IEEE Internet of Things Journal}, vol.~9, no.~9, pp. 6623--6637, 2022.

\bibitem{zhou2020dynamic}
J.~Zhou, W.~Zhao, and S.~Chen, ``Dynamic network slice scaling assisted by prediction in {5G} network,'' \emph{IEEE Access}, vol.~8, pp. 133\,700--133\,712, 2020.

\bibitem{chen2022designing}
H.-M. Chen, S.-Y. Chen, S.-K. Wang, and J.-Q. Chen, ``Designing a reinforcement learning approach for the {NFV} orchestration system with energy saving optimization,'' in \emph{2022 8th International Conference on Applied System Innovation (ICASI)}, 2022, pp. 98--101.

\bibitem{aba_ns_orchestrator_2024}
M.~A. Aba, M.~Kassis, M.~Elkael, A.~Araldo, A.~A. Khansa, H.~Castel-Taleb, and B.~Jouaber, ``Efficient network slicing orchestrator for {5G} networks using a genetic algorithm-based scheduler with kubernetes: Experimental insights,'' in \emph{2024 IEEE 10th International Conference on Network Softwarization (NetSoft)}, 2024, pp. 82--90.

\bibitem{xing_joint_2022}
C.~Xing, Y.~L. Lee, and Y.~C. Chang, ``Joint {Baseband} and {Radio} {Resource} {Allocation} for {5G} {Network} {Slicing} in {H}-{CRANs},'' in \emph{2022 {IEEE} {International} {Conference} on {Communications} {Workshops} ({ICC} {Workshops})}, May 2022, pp. 891--896, iSSN: 2694-2941.

\bibitem{miao_ddqn_based_2022}
J.~Miao, X.~Chai, X.~Song, and T.~Song, ``A {DDQN}-based {Energy}-{Efficient} {Resource} {Allocation} {Scheme} for {Low}-{Latency} {V2V} communication,'' in \emph{2022 {IEEE} 5th {International} {Electrical} and {Energy} {Conference} ({CIEEC})}, May 2022, pp. 53--58.

\bibitem{jayanthi_user_association_2021}
S.~S. Jayanthi, Y.~L. Lee, and Y.~C. Chang, ``User association for multi-tenant heterogeneous network slicing using genetic algorithm,'' in \emph{2021 8th International Conference on Computer and Communication Engineering (ICCCE)}, 2021, pp. 326--330.

\bibitem{ee_communication_tun_20}
Y.~K. Tun, D.~H. Kim, M.~Alsenwi, N.~H. Tran, Z.~Han, and C.~S. Hong, ``Energy efficient communication and computation resource slicing for {eMBB} and {URLLC} coexistence in {5G} and beyond,'' \emph{IEEE Access}, vol.~8, pp. 136\,024--136\,035, 2020.

\bibitem{Laroui_IWCMC_2020}
M.~Laroui, M.~A. Cherif, H.~I. Khedher, H.~Moungla, and H.~Afifi, ``Scalable and cost efficient resource allocation algorithms using deep reinforcement learning,'' in \emph{2020 International Wireless Communications and Mobile Computing (IWCMC)}, 2020, pp. 946--951.

\bibitem{Ravindran_EESO_2020}
S.~Ravindran, S.~Chaudhuri, J.~Bapat, and D.~Das, ``Eeso: Energy efficient system-resource optimization of multi-sub-slice-connected user in {5G RAN},'' in \emph{2020 IEEE International Conference on Electronics, Computing and Communication Technologies (CONECCT)}, 2020, pp. 1--6.

\bibitem{zhang2023energy}
Y.~Zhang, L.~Zhao, K.~Liang, G.~Zheng, and K.-C. Chen, ``Energy efficiency and delay optimization of virtual slicing of fog radio access network,'' \emph{IEEE Internet of Things Journal}, vol.~10, no.~3, pp. 2297--2313, 2023.

\bibitem{tran_duc_dung_2024}
D.-D. Tran, V.~N. Ha, S.~K. Sharma, T.~T. Nguyen, S.~Chatzinotas, and P.~Popovski, ``Energy-efficient noma for 5g heterogeneous services: A joint optimization and deep reinforcement learning approach,'' \emph{IEEE Transactions on Communications}, pp. 1--1, 2024.

\bibitem{dig_twin_2024}
Z.~Zhang, Y.~Huang, C.~Zhang, Q.~Zheng, L.~Yang, and X.~You, ``Digital twin-enhanced deep reinforcement learning for resource management in networks slicing,'' \emph{IEEE Transactions on Communications}, vol.~72, no.~10, pp. 6209--6224, 2024.

\bibitem{dig_twin_6g_2024}
\BIBentryALTinterwordspacing
M.~Yaqoob, R.~Trestian, M.~Tatipamula, and H.~X. Nguyen, ``Digital-twin-driven end-to-end network slicing toward {6G},'' \emph{IEEE Internet Computing}, vol.~28, no.~2, p. 47–55, Mar. 2024. [Online]. Available: \url{https://doi.org/10.1109/MIC.2023.3332252}
\BIBentrySTDinterwordspacing

\bibitem{dig_twin_5g_adm_control_2024}
\BIBentryALTinterwordspacing
J.~Wang, J.~Li, and J.~Liu, ``Digital twin-assisted flexible slice admission control for {5G} core network: A deep reinforcement learning approach,'' \emph{Future Gener. Comput. Syst.}, vol. 153, no.~C, p. 467–476, Apr. 2024. [Online]. Available: \url{https://doi.org/10.1016/j.future.2023.12.018}
\BIBentrySTDinterwordspacing

\end{thebibliography}
\endgroup

\section*{BIOGRAPHY}



 


\begin{IEEEbiography}
    [{\includegraphics[width=1in,height=1.25in,clip,keepaspectratio]{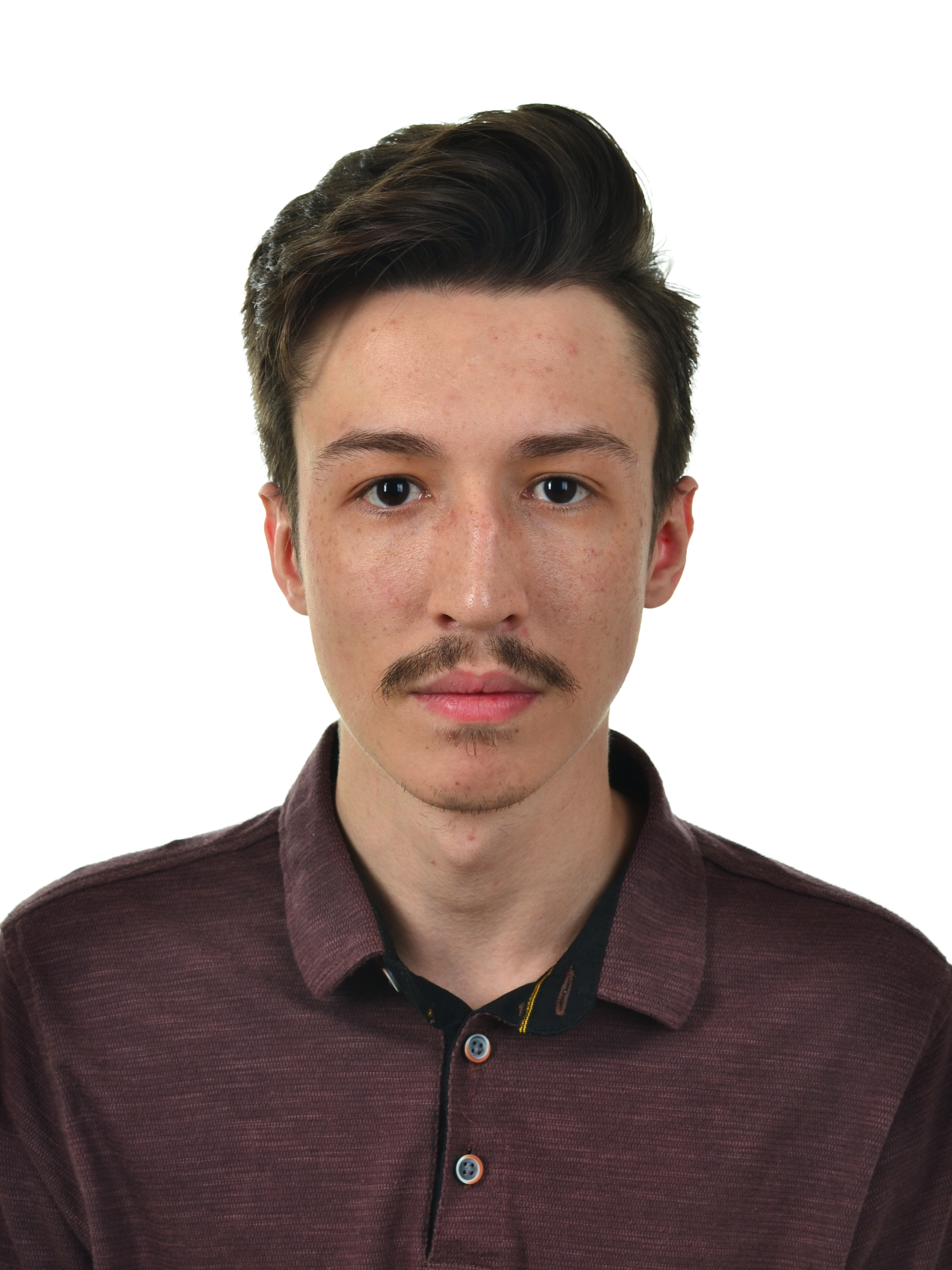}}]{\bf{Adnei W. Donatti}}
    is a Ph.D. student in Computer Engineering at the Escola Politécnica of the University of São Paulo (USP), under the Graduate Program in Electrical Engineering (PPGEE). He conducts research on energy-efficient Network Slicing with a focus on intelligent Virtual Network Function (VNF) placement and orchestration, leveraging Deep Reinforcement Learning (DRL) techniques. His recent work explores scalable, cross-layer optimization frameworks for 5G and beyond, integrating AI-based decision systems and sustainability-aware policies. He is actively involved in the SFI² (Smart Future Internet Infrastructure) project and leads the development of reproducible pipelines for large-scale genotype and phenotype data analysis using Spark and R on cloud platforms. He holds an M.Sc. in Applied Computing and a B.Tech. in Information Technology from Santa Catarina State University (UDESC). His research interests include network softwarization, cloud computing, distributed systems, and sustainable computing.
\end{IEEEbiography}


\begin{IEEEbiography}
[{\includegraphics[width=1in,height=1.25in,clip,keepaspectratio]{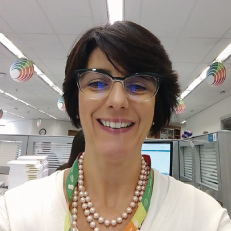}}]{Marcia Cristina Machado}
Graduated in Computer Science from the University of Guarulhos (1994), Master’s in Information Systems from the Pontifical Catholic University of São Paulo (2002), Master's in Business Administration from University Nove de Julho (2016), Ph.D. in Computer Engineering and Digital Systems from the University of São Paulo. She is currently a PosDoc at the Polytechnic School of the University of São Paulo, carrying out research on energy efficiency, sustainability indicators, and machine learning applications. She works as a guest professor on MBA courses at POLI-USP and FIA-FEA / USP.
\end{IEEEbiography}


\begin{IEEEbiography}
[{\includegraphics[width=1in,height=1.25in,clip,keepaspectratio]{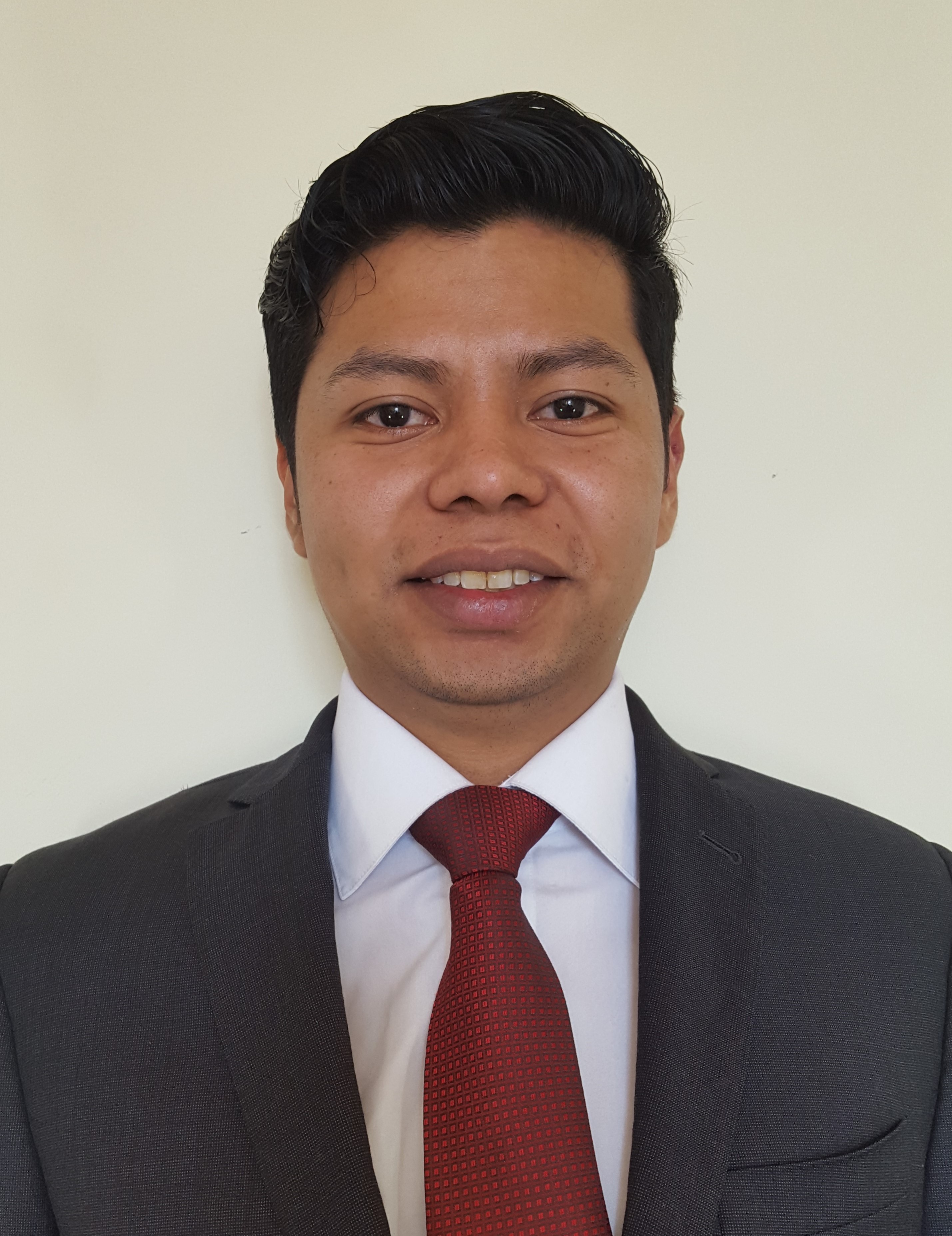}}]{Marvin A. Lopez Martine}
Ph.D. student at the Federal University of Pernambuco (UFPE) working on power consumption of Network Slicing in the SFI2 project. M.Sc. degree in Information Technologies from Nikola Vaptsarov Naval Academy (NVNA), and a Bachelor's degree in International Business from the Technological University of El Salvador (UTEC). Interested in slice energy consumption, slice energy efficiency, and energy consumption modeling.
\end{IEEEbiography}


\begin{IEEEbiography}
[{\includegraphics[width=1in,height=1.25in,clip,keepaspectratio]{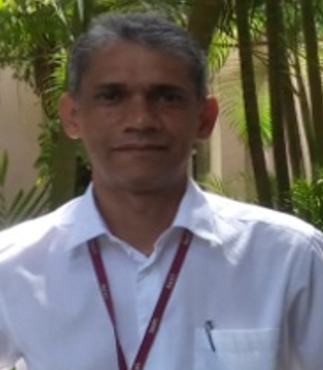}}]{Sabino Rogério S. Antunes}
Ph.D. student at the Federal University of Pernambuco (UFPE). M.Sc. degree in Computer Science from the Federal University of Pernambuco - UFPE (2018). Universidade Salgado de Oliveira - UNIVERSO (2006), with a specialization in Network Systems Security from Faculdade Santa Maria – FSM (2011). Worked as a lecturer at Universidade Joaquim Nabuco in the Computer Networks course (2011 – 2019). Currently holds the position of IT technician at UFPE. Interested in Cloud Computing, Computer Networks, Mobile Networks, and Network Slicing.
\end{IEEEbiography}


\begin{IEEEbiography}
[{\includegraphics[width=1in,height=1.25in,clip,keepaspectratio]{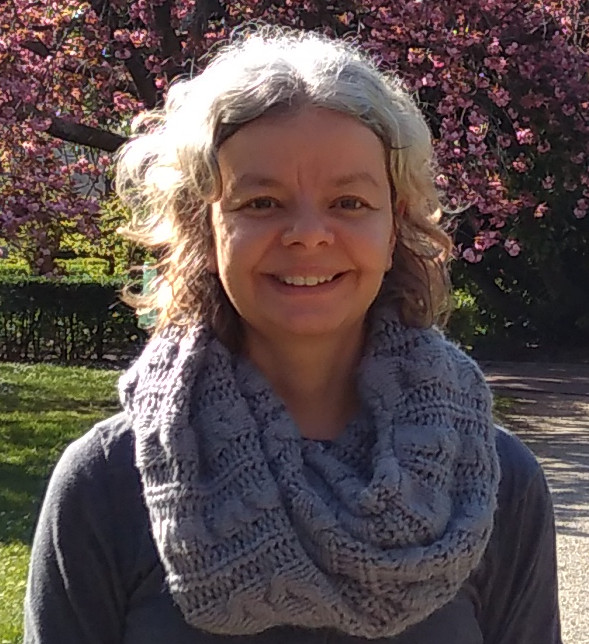}}]{\bf{Sand L. Corrêa} -}
Sand L. Corrêa received a bachelor's degree in Computer Science from the Federal University of Goiás (UFG), in 1994. In 1997, she received an M.Sc. degree in Computer Science from the State University of Campinas (Unicamp). She received a D.Sc. degree in Informatics from the Pontifical Catholic University of Rio de Janeiro (PUC-Rio), in 2011. Since 2010, she is an associate professor at the Institute of Informatics at UFG.
\end{IEEEbiography}


\begin{IEEEbiography}
[{\includegraphics[width=1in,height=1.25in,clip,keepaspectratio]{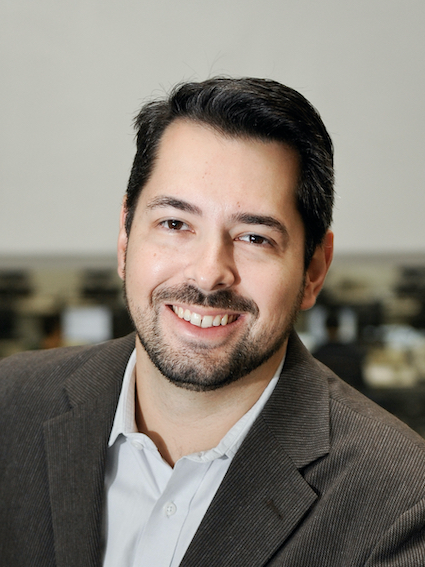}}]{\bf{Tiago C. Ferreto}} is an associate professor of Computer Science at the Pontifical Catholic University of Rio Grande do Sul (PUCRS), Brazil. He received his Ph.D. in Computer Science from the Computer Science Department, PUCRS, Brazil (2010) with a Sandwich Ph.D. Internship at TU-Berlin, Germany (2007-2008). He is currently the head of GRIN (Research Group on Networking, Infrastructure, and Cloud Computing) and LAD (High-Performance Computing Laboratory) at PUCRS. His research interests include Cloud and Edge Computing, IT Infrastructure Management, Computer Networks, High-Performance Computing, and Big Data.
\end{IEEEbiography}


\begin{IEEEbiography}
[{\includegraphics[height=1.25in,clip,keepaspectratio]{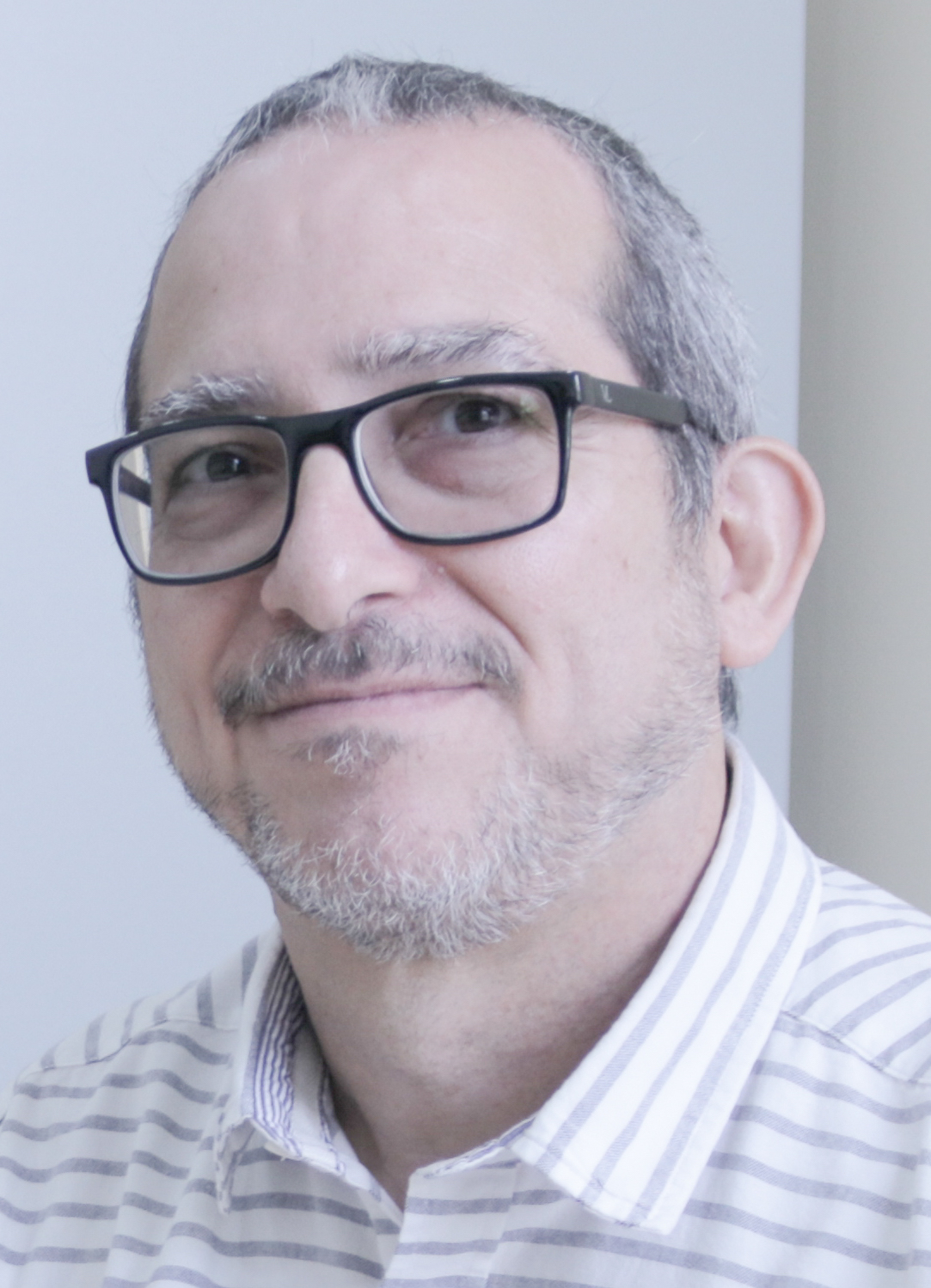}}]{\bf{José A. Suruagy}  (Life Senior Member, IEEE) -} received his Ph.D. in Computer Science from the University of California at Los Angeles -- UCLA (1990). Has experience in Computer Science, focusing on Telecomputing, acting on the following subjects: computer networks, measurement, quality of service, performance evaluation, and cybersecurity. In 2014, received the “Computer Network Brazilian Symposium Distinguished Award” for his scientific contributions in the areas of computer networks and distributed systems, for his participation in the Symposium activities, and for his services that benefited that community. In 2019, he received the “Newton Faller Award” from the Brazilian Computer Society, in recognition of his brilliant collaboration and dedication to the exercise of multiple activities of great interest to the Brazilian Computing Community, in particular, his work with the Forum of Computer Science Graduate Chairs. Currently, he is a professor and researcher at CESAR.School.
\end{IEEEbiography}


\begin{IEEEbiography}
[{\includegraphics[width=1in,height=1.25in,clip,keepaspectratio]{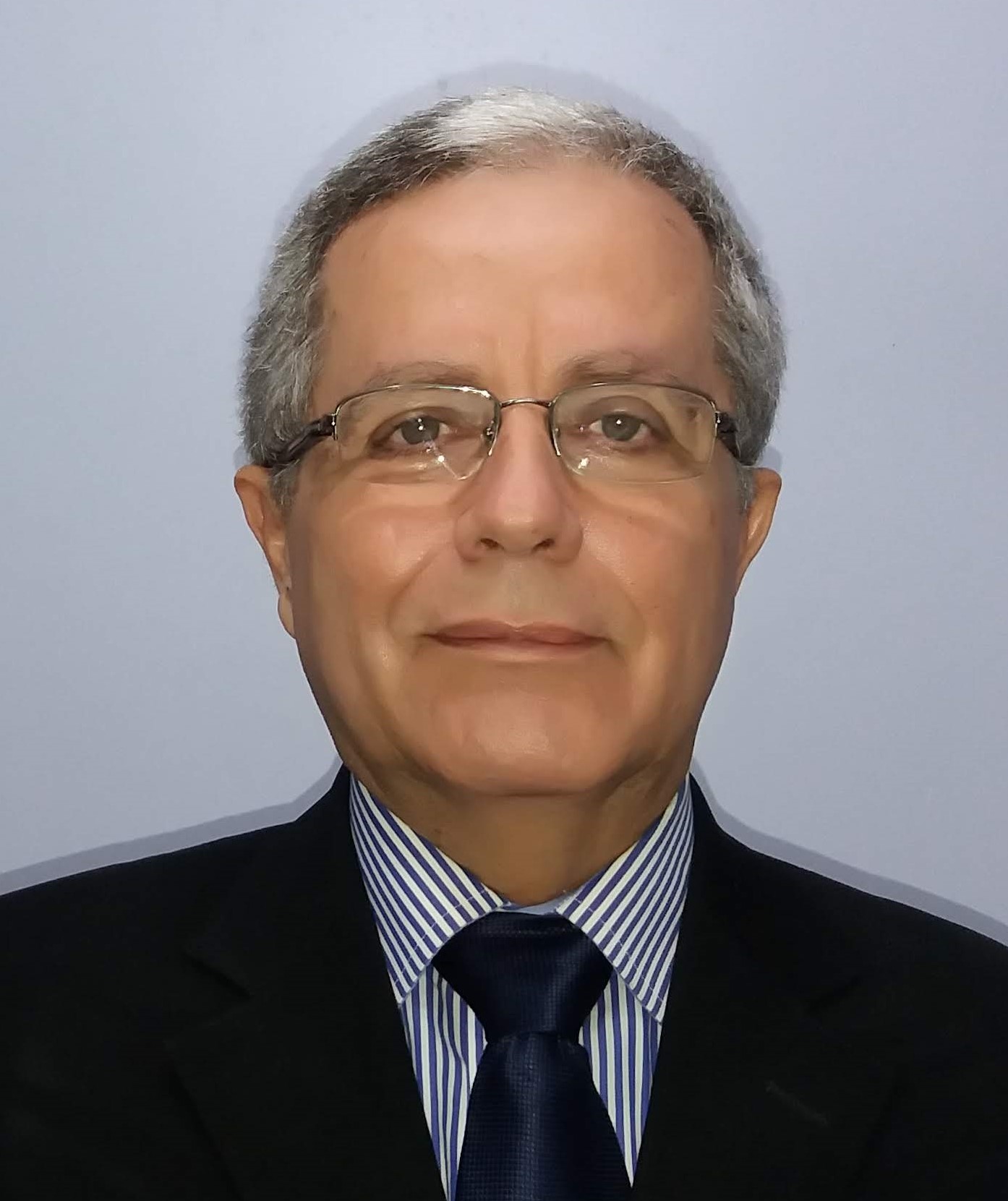}}]{\bf{Joberto S. B. Martins} (Life Senior Member, IEEE) -} Ph.D. in Computer Science at Université Pierre et Marie Curie - UPMC, Paris (1986), PosDoc at ICSI - Berkeley University (1995), and PosDoc Senior Researcher at Paris Saclay University - France (2016). International Professor at Hochschule für Technik und Wirtschaft des Saarlandes - HTW (Germany) (since 2004). Full Professor at Salvador University (UNIFACS) on Computer Science, Director of NUPERC research group with research interests in Network Slicing, Resource Orchestration, Machine Learning, and Smart City. He is a key speaker, teacher, and invited lecturer at various international congresses and companies in Brazil and Europe.
\end{IEEEbiography}


\begin{IEEEbiography}
[{\includegraphics[width=1in,height=1.25in,clip,keepaspectratio]{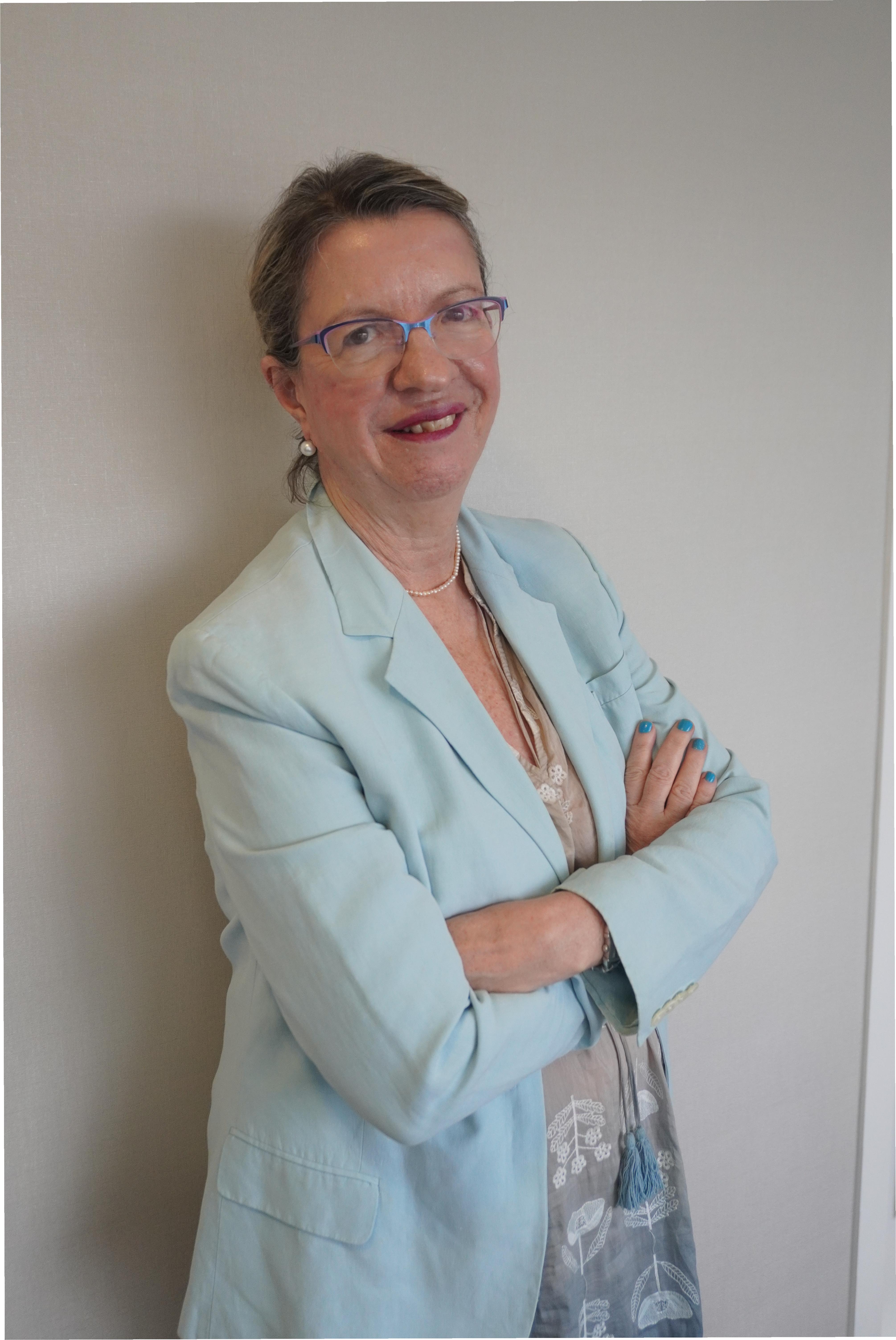}}]{\bf{Tereza C. Carvalho} -} Associate Professor of Escola Politécnica – University of São Paulo (USP) and visiting professor at Université Paris 1 Panthéon-Sorbonne. She is the founder and general coordinator of LASSU (Laboratory of Sustainability on ITC) since 2010 and CEDIR-USP (Center for Reuse and Discard of Informatics Residuals) since 2009. She is a former assessor of CTI – USP (Information Technology Coordination) from 2010-2013 and CCE-USP (Electronic Computing Center) director from 2006-2010. She is a Sloan Fellow 2002 from MIT (Massachusetts Institute of Technology). She has coordinated international and national R\&D projects since 2000 in Green Computing, Cloud Computing, IT Energy Efficiency, IT Governance, Digital Technologies applied to the Amazon Production Chains, WEEE (Waste Electrical and Electronic Equipment), Future Internet, Scientific DMZ, and Security. She holds several international patents.
\end{IEEEbiography}


\vfill

\EOD

\end{document}